%% file: main.tex
\documentclass[reprint, amsmath,amssymb,aps,pre]{revtex4-2}
\usepackage{graphicx,color}
\usepackage{import}
\usepackage{dcolumn}
\usepackage{bm}
\begin{document}
\title{Intrinsic Langevin dynamics of rigid inclusions on curved surfaces}
\author{Balázs Németh}
\email{bn273@cam.ac.uk}
\author{Ronojoy Adhikari}
\email{ra413@cam.ac.uk}
\affiliation{Department of Applied Mathematics and Theoretical Physics, Centre
for Mathematical Sciences, University of Cambridge, Wilberforce Road,
Cambridge CB3 0WA, United Kingdom}
\date{\today}
\begin{abstract}
    The stochastic dynamics of a rigid inclusion constrained to move on a curved surface has many applications in biological and soft matter physics, ranging from the diffusion of passive or active membrane proteins to the motion of phoretic particles on liquid-liquid interfaces. Here we construct intrinsic Langevin equations for an oriented rigid inclusion on a curved surface using Cartan's method of moving frames. We first derive the Hamiltonian equations of motion for the translational and rotational momenta in the body frame. Surprisingly, surface curvature couples the linear and angular momenta of the inclusion. We then add to the Hamiltonian equations linear friction, white noise and arbitrary configuration-dependent forces and torques to obtain intrinsic Langevin equations of motion in phase space. We provide the integrability conditions, made non-trivial by surface curvature, for the forces and torques to admit a potential, thus distinguishing between passive and active stochastic motion. We derive the corresponding Fokker-Planck equation in geometric form and obtain fluctuation-dissipation relations that ensure  Gibbsian equilibrium. We extract the overdamped equations of motion by adiabatically eliminating the momenta from the Fokker-Planck equation, showing how a peculiar cancellation leads to the naively expected Smoluchowski limit. The overdamped equations can be used for accurate and efficient intrinsic Brownian dynamics simulations of passive, driven and active diffusion processes on curved surfaces. Our work generalises to the collective dynamics of many inclusions on curved surfaces.
\end{abstract}
\maketitle

\section{Introduction}

The stochastic dynamics of a rigid inclusion on a curved surface provides an useful theoretical  model for quantitatively describing several important experimental phenomena in biological and soft matter physics, for instance the diffusion of passive \cite{Peters1982,Cone1972} or active \cite{Noji1997} proteins on the cell membrane or the motion of phoretic particles confined to liquid-liquid interfaces \cite{Grzybowski2000,Grzybowski2001,Fei2018}. In these situations, the inertia of the particle is typically negligible and a theoretical description in terms of overdamped equations of motion is appropriate. Accordingly, the existing literature on the stochastic motion of particles on curved surfaces has focussed on the overdamped limit \cite{Sknepnek2015,Fily2016,Apaza2018,CastroVillarreal2018}. There are, however, at least three separate mathematically subtle issues in describing the generic stochastic motion of inclusions on curved surfaces. 

The first issue in formulating dynamics is that local coordinates have no intrinsic meaning on manifolds. Therefore, the physical content of the equations of motion must be expressed in a manner that is invariant under change of coordinates. This poses a problem that is peculiar to stochastic processes, since there are at least two prescriptions for how the change of variables should be implemented in the stochastic equations of motion. This is the well-known Ito-Stratonovich dilemma \cite{vanKampen1981,Gardiner2004}.

For the specific case of the motion of a rigid inclusion on a curved surface, the natural description of the motion is in terms of a body frame aligned with the principal axes of inertia and the position of the inclusion on the surface. Therefore, there is a natural basis, provided by the body frame, in which to express tensor fields on the manifolds. This is related to Cartan's method of moving frames \cite{Cartan1935,Sharpe1997,Clelland2017} with the crucial difference being that the frame represents a physical degree of freedom and does not have any redundant "gauge" degrees of freedom usually appearing in the classical applications of the method. This leads to a formulation of the equations of motion using moving frames, giving, as we shall show below, a clear conceptual picture of the kinematics of an inclusion on a curved surface. 

The second issue is that the overdamped limit is a singular limit, in the sense that the highest time derivative in the problem is multiplied by a small parameter which is eventually taken to zero \cite{Hinch1991}. For stochastic equations of motion, this process of taking the singular limit usually introduces additional terms (often called noise-induced spurious drift \cite{Volpe2010,Birrell2016}), unanticipated by a naive realisation of the limit.

Therefore, it is essential to start from the inertial equations of motion, which remain unambiguous if linear coordinate-dependent friction is present, and systematically derive the overdamped limit by an adiabatic elimination procedure \cite{Murphy1972,Wilemski1976,Gardiner1984}. This process can be expected to have additional subtleties on curved spaces as geometric quantities associated with the surface, appearing in the inertial equations of motion can remain in the overdamped limit \cite{Birrell2016}. To formulate the intrinsic Langevin equations and to derive the overdamped limit carefully is one of the main purposes of our work.

A third issue, of great contemporary interest, is identifying forces and torques that do not derive from a potential. These are termed active forces and torques in the contemporary literature, but have also been referred to previously as circulatory forces \cite{Ziegler1977}. Therefore, we need to identify when force-torque pairs, given as a function of the body frame coordinates, derive (or do not) from a potential.

Obtaining the integrability conditions in our setting is non-trivial because the body frame velocities are non-holonomic, i.e. they are not time-derivatives of coordinate functions. The integrability conditions, therefore, have a non-trivial character which we elucidate using ideas from exterior calculus \cite{Frankel2011}.

The remainder of the paper is organised as follows. In Section II, we first construct the kinematic equations of motion, in which the non-holonomic body frame components of the velocity are used in preference to holonomic velocities. It is natural to use non-holonomic velocities in the moving frame method in which a preference is given to the simplicity of the kinematic description over the choice of a coordinate basis. We then construct the dynamical equations of motion in the Lagrangian picture for diffusive motion in a potential landscape on the surface. These are transformed into the Hamiltonian picture with non-canonical momenta and the equations of motion are expressed in terms of Poisson brackets. To these Poisson bracket equations we add linear friction, white noise and forces and torques of a general character. We identify conditions for when the latter derive from a potential. In Section III, having established the pathwise equations of motion, we focus on the equation of motion for the probability densities and derive the Fokker-Planck equation in phase space, expressing it in geometric form. From this we obtain the fluctuation-dissipation relation when the forces and torques derive from a potential. In Section IV we perform a systematic adiabatic elimination procedure for the momenta in the case of large friction, and obtain the Smoluchowski equation for the probability density in configuration space. Identifying the corresponding stochastic equations of motion suggests a reliable intrinsic method for performing Brownian dynamics simulations. Finally, in Section V we present a detailed application of our general formalism to the case of a rigid inclusion on a sphere, and we draw conclusions in Section VI. 
\section{Equations of motion}
\subsection{Method of moving frames}\label{subsec:geom_prelim}
\begin{figure}[ht]
	\centering
	\def\svgwidth{\columnwidth}
	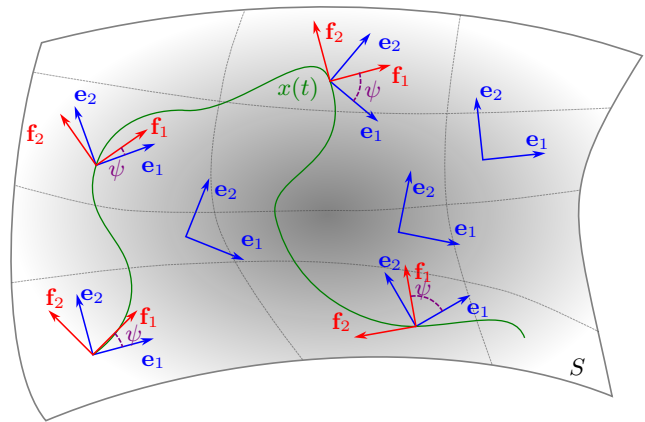
	\caption{Illustration of moving frames and coordinate systems on the surface $S$.}
	\label{fig:moving_frame}
\end{figure}
In this section we establish some notation and recall useful concepts from the classical differential geometry of surfaces via the method of moving frames \cite{Carmo1994,Flanders1989}.

Let $S$ be a smooth embedded oriented surface in $\mathbb{R}^3$. The orthonormal frame bundle $FS$ of the surface $S$ is the set of tuples of the form $\left(x,\mathbf{f}_{1},\mathbf{f}_{2}\right)$, where $x\in S$ and $\left\{ \mathbf{f}_{1},\mathbf{f}_{2}\right\}$ is a right-handed orthonormal basis of the tangent space $T_{x}S$. It is a smooth manifold of dimension three, on which we can choose local coordinates as follows. We first pick a local coordinate system $x^\alpha$ on $S$, with the index $\alpha$ running from $1$ to $2$ as well as a local orthonormal frame field $\mathbf{e}_a$ on $S$, which can be done by e.g. performing the Gram-Schmidt process on the coordinate basis vectors $\frac{\partial}{\partial x^\alpha}$. Now if $q=\left(x,\mathbf{f}_a\right)\in FS$ is a point in the domain of the local coordinate chart and frame field, we can parametrize $x$ by $x^\alpha$ and a unique angle $\psi\in [0,2\pi)$ such that
\begin{equation}
	\begin{pmatrix}
		\mathbf{f}_1 & \mathbf{f}_2
	\end{pmatrix}=\begin{pmatrix}
		\mathbf{e}_1 & \mathbf{e}_2
	\end{pmatrix}\begin{pmatrix}
		\cos\psi & -\sin\psi\\
		\sin \psi & \cos \psi
	\end{pmatrix}\label{eq:mov_frame_def}.
\end{equation}
Thus the triplet $\left(x^1,x^2,\psi\right)$ forms a valid local coordinate system on $FS$. It is also useful to introduce the dual basis $\theta^a$ of one-forms to $\mathbf{e}_a$, defined by the relations
\begin{equation}
	\theta^a\left(\mathbf{e}_b\right)=\delta^a_b.
\end{equation}
The one-forms $\theta^a$ are also referred to as an orthonormal coframe field since they are orthonormal with respect to the inverse metric. In a local coordinate system we may expand both the frame and coframe fields as
\begin{eqnarray}
	\mathbf{e}_a&=&e_a^\alpha\frac{\partial}{\partial x^\alpha},\\
	\theta^a&=&\theta^a_\alpha dx^\alpha.
\end{eqnarray}
The coframe fields induce a Riemannian area element $\mathrm{vol}_S$ on the surface as:
\begin{equation}
	\mathrm{vol}_S=\theta^1\wedge\theta^2=\sqrt{\det g}dx^1\wedge dx^2.
\end{equation}

The Riemannian manifold $S$ can be equipped with a distinguished connection, the Levi-Civita connection, defined as the unique metric-compatible torsion-free connection. With respect to the local frame field $\mathbf{e}_a$, it can be described by a connection one-form $\Gamma=\Gamma_\alpha dx^\alpha$ as follows:
\begin{eqnarray}
	D\mathbf{e}_1&=&\mathbf{e}_2\Gamma,\label{eq:conn_def_1}\\
	D\mathbf{e}_2&=&-\mathbf{e}_1\Gamma,\label{eq:conn_def_2}
\end{eqnarray}
where $D\mathbf u$ denotes the covariant derivative of the vector field $\mathbf u$. It is important to note that both the coordinate $\psi$ and the connection form $\Gamma$ depends on the arbitrary choice of moving frame $\mathbf{e}_a$. In the physics literature, this choice is called a gauge, and a change of the moving frame is a gauge transformation. Under this transformation, the connection form $\Gamma$ changes to $\Gamma'=\Gamma+d\varphi$, where $\varphi(x)$ is an arbitrary smooth real-valued function representing the angle between the new and the old gauge. Notice, however, that the two-form
\begin{equation}
	d\Gamma'=d\Gamma=-\kappa(x)\mathrm{vol}_S\label{eq:curv_conn_form}
\end{equation}
is unchanged, since the exterior derivative of the exact one-form $d\varphi$ vanishes. As the space of two-forms over $S$ is one-dimensional, $d\Gamma$ has to be proportional to the Riemannian area form, and the constant of proportionality —representing the gauge-invariant curvature of the connection — is an intrinsic geometric quantity that is in fact equal to minus the Gaussian curvature $\kappa(x)$ of the surface at the point $x$ \cite{Carmo1994}.

The connection can be used to define the parallel transport of frames on $FS$. Let $q=\left(x,\mathbf{f}_1,\mathbf{f}_2\right)\in FS$ be an orthonormal frame. The infinitesimal generator of parallel transport along the frame vector $\mathbf{f}_a$ is a vector field $V_a$ on $FS$, whose coordinate expression is given by
\begin{equation}
	V_a=f_a^\alpha\left(\frac{\partial}{\partial x^\alpha}-\Gamma_\alpha\frac{\partial}{\partial\psi}\right).\label{eq:hor_vec}
\end{equation}
Equation \eqref{eq:hor_vec} follows from the fact that infinitesimal parallel transport along $\mathbf{f}_a$ shifts the basepoint $x$ of the frame by the vector $\mathbf{f}_a=f_a^\alpha\frac{\partial}{\partial x^\alpha}$ while rotating the frame by an amount $-\Gamma\left(\mathbf{f}_a\right)=-f_a^\alpha\Gamma_\alpha$ so that the frame itself remains covariantly constant, i.e. parallel. The vector field generating infinitesimal rotations of the frame is simply
\begin{equation}
	V_R=\frac{\partial}{\partial\psi}\label{eq:vert_vec}.
\end{equation}
In the mathematical literature, the vector fields $V_a$ are called horizontal vector fields, while $V_R$ is a vertical vector field \cite{Sharpe1997,Frankel2011}. In the sequel, we are going to use the notation $V_i$ for the collection of horizontal and vertical vector fields, with the convention that the index $i$ runs from $1$ to $3$ and $V_3\equiv V_R$.

Since the vectors $V_i$ form a global basis of vector fields for each tangent space of $FS$, we can also construct the corresponding global basis of one-forms $\eta^i$ defined by the relations
\begin{equation}
	\eta^i\left(V_j\right)=\delta^i_j.
\end{equation}In particular, let
\begin{equation}
	\begin{pmatrix}
		\Theta^1\\
		\Theta^2
	\end{pmatrix}\triangleq\begin{pmatrix}
		\cos\psi&\sin\psi\\
		-\sin\psi&\cos\psi
	\end{pmatrix}\begin{pmatrix}
		\theta^1\\
		\theta^2
	\end{pmatrix}
\end{equation}be the dual coframe basis to $\mathbf{f}_a$, we have:\begin{eqnarray}
	\eta^a&=&\Theta^a=\Theta^a_\alpha dx^\alpha\label{eq:coftr}\\
	\eta^3&=&\Gamma+d\psi=\Gamma_\alpha dx^\alpha+d\label{eq:cofrot}\psi
\end{eqnarray}
\subsection{Kinematics}
\begin{figure}[ht]
    \centering
    \def\svgwidth{\columnwidth}
    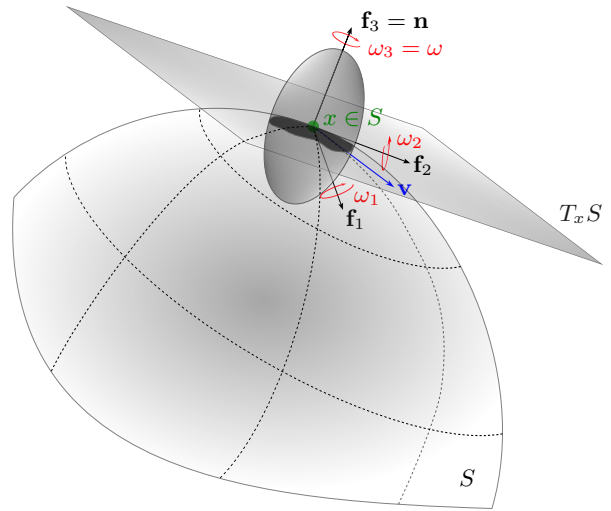
    \caption{Representation of an inclusion on a curved surface $S$.}
    \label{fig:constr_motion}
\end{figure}
We model the inclusion as a rigid body with both its positional and orientational degrees of freedom constrained. We assume that the centre of mass of the inclusion is constrained to lie on a smooth embedded oriented surface $S\subset\mathbb R^3$, while one of its principal axes of inertia $\mathbf{f}_3$ is constrained to be normal to the surface. Therefore the configuration the inclusion at any instance of time can be described by the tuple $\left(x,\mathbf{f}_{1},\mathbf{f}_{2}\right)$, where $x\in S$ is the position of its centre of mass and $\left\{\mathbf{f}_1,\mathbf{f}_2\right\}$ are the two other principal axes of inertia, forming a right-handed orthonormal basis of the tangent plane $T_xS$. As a result, the configuration space of the inclusion is precisely the orthonormal frame bundle $FS$ of the surface.

The fundamental kinematical equations of motion are:
\begin{eqnarray}
    dx&=&\mathbf{f}_Iv^Idt,\label{eq:tr_kin}\\
    d\mathbf{f}_I&=&\mathbf{f}_J\omega^J_Idt,\quad I=1,2,3,\label{eq:rot_kin}
\end{eqnarray}
where $v^I(t)$ are the components of the translational velocity of the body in the body frame, while $\omega^I_J(t)$ are the components of the angular velocity of the body, represented by a $3\times 3$ antisymmetric matrix, also with respect to the body frame. (Uppercase indices $I,J$ run from $1$ to $3$ and repeated indices are summed over.) The matrix $\omega_I^J(t)$ is related to the usual angular velocity vector $\left(\omega_1\quad \omega_2\quad \omega_3\right)^T$ via
\begin{equation}
    \omega^I_J=\begin{pmatrix}
        0&-\omega_3&\omega_2\\
        \omega_3&0&-\omega_1\\
        -\omega_2&\omega_1&0
    \end{pmatrix}.\label{eq:om_matr_def}
\end{equation}

Since the curve $x(t)$ lies on $S$, the inclusion cannot have any normal velocity component to the surface. Furthermore, the rate of change of the normal must be tangential to the surface, giving, by the Weingarten equation,
\begin{equation}
    d\mathbf{f}_3=-s^a_b\mathbf{f}_a v^bdt\label{eq:weingarten},
\end{equation}
where $s^a_b(x)$ are the components of the second fundamental form of the surface at $x$ with respect to the basis $\mathbf{f}_a$ of the local tangent space. (Lowercase indices $a,b$ run from $1$ to $2$, they are raised and lowered with the standard metric $\delta_{ab}$ and repeated indices are summed over.) Combining \eqref{eq:weingarten} and \eqref{eq:rot_kin}, the kinematical constraints may be encapsulated in the following algebraic form:
\begin{eqnarray}
    v^3&=&0\label{eq:constr_tr},\\
    \omega^a_3&=&-s^a_bv^b\label{eq:constr_rot}.
\end{eqnarray}
It follows that the kinematics is fully specified by the pair of translational velocity components $v^1(t),v^2(t)$ and the angular velocity component $\omega(t)\triangleq \omega_3(t)$ normal to the surface.

The concepts introduced in the previous section allow us to connect the angular velocity of the inclusion to the covariant derivative of the its associated body frame as follows. By taking the dot product of \eqref{eq:rot_kin} for $I=1$ with $\mathbf{f}_2$ and using \eqref{eq:mov_frame_def} together with \eqref{eq:conn_def_1} we find 
\begin{eqnarray}
    \omega(t)&=&\mathbf{f}_2\cdot\frac{d\mathbf{f}_1}{dt}\nonumber\\
    &=&\frac{d\psi}{dt}+\mathbf{e}_2\cdot\frac{d\mathbf{e}_1}{dt}\nonumber\\
    &=&\frac{d\psi}{dt}+\mathbf{e}_2\cdot D_{\mathbf{v}}\mathbf{e}_1\nonumber\\
    &=&\frac{d\psi}{dt}+\Gamma(\mathbf{v})\nonumber\\
    &=&\frac{d\psi}{dt}+v^a\Gamma_\alpha f_a^\alpha.\label{eq:ang_vel_intr}
\end{eqnarray}
Equation \eqref{eq:ang_vel_intr} provides another useful interpretation of $\omega(t)$: it is the scalar-valued orthogonal projection to $S$ of the rate of change of the vector field $\mathbf{f}_1(t)$ along the curve $x(t)$. One of the fundamental results in the differential geometry of embedded surfaces in $\mathbb R^3$ states that this projection is independent of the embedding of the surface $S$ in ambient space and is given by the covariant derivative of the vector field $\mathbf{f}_1(t)$ along $x(t)$ with respect to the Levi-Civita connection of the Riemannian surface $S$ \cite{Carmo2016}. With the aid of \eqref{eq:ang_vel_intr}, we can project out the rate of change of surface normal to obtain the intrinsic form of the kinematical equations \eqref{eq:tr_kin}—\eqref{eq:rot_kin}:
\begin{eqnarray}
    dx&=&\mathbf{f}_av^adt,\label{eq:intr_kin_tr}\\
    D\mathbf{f}_a&=&\mathbf{f}_b\varepsilon^b_a\omega dt,\label{eq:intr_kin_rot}
\end{eqnarray}
where
\begin{equation}
    \varepsilon^a_b=\begin{pmatrix}
        0&-1\\
        1&0
    \end{pmatrix}.
\end{equation}
Therefore using \eqref{eq:intr_kin_rot} we can interpret $\omega$ as the covariant rate of rotation of the frame $\left\{\mathbf{f}_1(t),\mathbf{f}_2(t)\right\}$. Finally, observe that if $q(t):\left[t_0,t_1\right]\to FS$ is the trajectory of the inclusion, then we have:
\begin{equation}
    \dot q(t)=v^i(t)V_i\in T_{q(t)}FS\label{eq:tang_vec_exp},
\end{equation}
where we have introduced $v^i$ as the components of the generalized velocity of the inclusion, with the convention $v^R\equiv \omega$. Thus we can interpret $v^i(t)$ as the components of the tangent vector $\dot q(t)$ expanded in the basis $V_i$ of horizontal and vertical vector fields.
\subsection{Lagrangian dynamics}
The equations of motion for the inclusion are derived from a variational principle. The standard Lagrangian is the sum of translational and rotational kinetic energies minus a potential energy
\begin{equation}
L=\frac12\sum_{a=1}^2m\left(v^a\right)^2+\frac12\sum_{l=1}^3\mathbb{I}_l\left(\omega_l\right)^2-U(q)\label{eq:lagr},
\end{equation}
where $m$ is the mass of the inclusion and $\mathbb{I}_l$ are its principal moments of inertia. (Without loss of generality we may assume that the $\mathbf{f}_a$-s are aligned with the principal axes of inertia). For simplicity, we are going to assume that $\mathbb{I}_1=\mathbb{I}_2\triangleq \mathbb{I}$ and use the notation $J\triangleq \mathbb{I}_3$ for the moment of inertia along the surface normal. Substituting \eqref{eq:constr_rot} in \eqref{eq:lagr} yields, using the summation convention:
\begin{equation}L=\frac12\left(m\delta_{ab}+\mathbb{I}\delta_{cd}s^c_as^d_b\right)v^av^b+\frac12J\omega^2-U(q).\label{eq:lagr2}
\end{equation}
The intutitive interpretation of \eqref{eq:lagr2} is that the inclusion obtains extra translational mass on top of $m$ because it has to rotate its orientation in order to stay normal to the surface. However, this contribution is generally small: if $l=\sqrt{J/m}$ is roughly the size of the inclusion and $k$ is the typical curvature of the surface (having units of inverse length), then we have
\begin{equation}
    \frac{\mathbb{I}\delta_{cd}s^c_as^d_b}{m}\sim\frac{ml^2k^2}{m}=(lk)^2\ll 1,
\end{equation}
since for our model to be valid the inclusion size has to be much smaller than the scale over which the manifold is curved. Therefore we may neglect the second term in \eqref{eq:lagr2} to obtain
\begin{equation}
    L=\frac12m\delta_{ab}v^av^b+\frac12J\omega^2-U(q).\label{eq:lagr_intr}
\end{equation}
A remarkable property of the Lagrangian \eqref{eq:lagr_intr} is that it is purely intrinsic, i.e. it does not involve any extrinsic quantity coming from the embedding of the surface $S$ in ambient space. It is also useful to have a local coordinate expression of \eqref{eq:lagr_intr}:
\begin{equation}
    L=\frac12mg_{\alpha\beta}\dot{x}^\alpha\dot{x}^\beta+\frac12J\left(\dot\psi+\Gamma_\alpha\dot{x}^\alpha\right)^2-U\left(x^\alpha,\psi\right).\label{eq:lagr_coord}
\end{equation}
From equation \eqref{eq:lagr_coord} we could write down the Euler-Lagrange equations directly in local coordinates. However, they would reveal little insight about the intrinsic geometric structure of the equations motion, therefore we turn to the Hamiltonian picture instead where the equations of motion can be derived and analysed elegantly using Poisson brackets.
\subsection{Hamiltonian dynamics}
In this subsection we derive the governing equations of motion in Hamiltonian form in the body frame using Poisson brackets. Let $q^\mu=\left(x^\alpha,\psi\right)$ be our set of generalized coordinates with corresponding generalized momenta $p_\mu=\left(p_\alpha,s\right)$, where the index $\mu$ runs from $1$ to $3$. We have:
\begin{eqnarray}
    p_\alpha&=\frac{\partial L}{\partial \dot{x}^\alpha}=&mg_{\alpha\beta}\dot{x}^\beta+J\left(\dot\psi+\Gamma_\beta\dot{x}^\beta\right)\Gamma_\alpha,\\
    s&=\frac{\partial L}{\partial \dot \psi}=&J\left(\dot\psi+\Gamma_\beta\dot{x}^\beta\right)=J\omega.\label{eq:body_ang_mom_def}
\end{eqnarray}
Performing a Legendre transformation, we find the Hamiltonian to be:
\begin{eqnarray}
    H=\frac{g^{\alpha\beta}\left(p_\alpha-s\Gamma_\alpha\right)\left(p_\beta-s\Gamma_\beta\right)}{2m}+\frac{s^2}{2J}+U(q)\label{eq:ham_coord}.
\end{eqnarray}
It is instructive to introduce the body frame linear momenta
\begin{equation}
    \pi_a=\frac{\partial L}{\partial v^a}=m\delta_{ab}v^b=mv_a.\label{eq:body_lin_mom_def}
\end{equation}
These are related to the canonical momenta as:
\begin{equation}
    \pi_a=f_a^\alpha\left(p_\alpha-s\Gamma_\alpha\right).\label{eq:body_mom_from_can}
\end{equation}
From now on, we will refer to the triplet $\left(\pi_1,\pi_2,s\right)$ with $\left(\pi_i\right)$, where the index $i$ running from $1$ to $3$ with the convention that $\pi_3=s$.

In terms of the body frame momenta, the Hamiltonian \eqref{eq:ham_coord} assumes the standard quadratic form
\begin{equation}
    H=\frac{\delta^{ab}\pi_a\pi_b}{2m}+\frac{s^2}{2J}+U(q)=\frac12G^{ij}\pi_i\pi_j+U(q),\label{eq:ham_quad}
\end{equation}
where $G^{ij}$ denotes the inverse of the inertia metric
\begin{equation}
    G_{ij}=\begin{pmatrix}
        m&0&0\\
        0&m&0\\
        0&0&J
    \end{pmatrix}.\label{eq:inertia_metric}
\end{equation}
The transformation \eqref{eq:body_mom_from_can} between the body frame momenta and the canonical momenta can be succinctly captured using the components $V_i=V_i^\mu(q)\frac{\partial}{\partial q^\mu}$ of the horizontal \eqref{eq:hor_vec} and vertical vector fields \eqref{eq:vert_vec} as:
\begin{equation}
    \pi_i=V_i^\mu(q)p_\mu.
\end{equation}
The phase space carries a canonical Poisson structure that reads in canonical coordinates $\left(q^\mu,p_\nu\right)$ as:
\begin{equation}
    \left\{q^\mu,q^\nu\right\}=\left\{p_\mu,p_\nu\right\}=0,\quad \left\{q^\mu,p_\nu\right\}=\delta^\mu_\nu.
\end{equation}
After some calculations detailed in the appendix, the Poisson structure in terms of the body frame momenta are given by:
\begin{eqnarray}
    \left\{\pi_1,\pi_2\right\}&=&-\kappa(x)s,\label{eq:pb1}\\
    \left\{\pi_1,s\right\}&=&\pi_2,\label{eq:pb2}\\
    \left\{\pi_2,s\right\}&=&-\pi_1.\label{eq:pb3}
\end{eqnarray}
Since these relations are linear in the momenta, we collect them as:\begin{equation}
    \left\{\pi_i,\pi_j\right\}=\Omega^k_{ij}(q)\pi_k.
\end{equation}
The Poisson brackets of the generalized coordinates and the body frame momenta are given by the components of the horizontal and vertical vector fields \eqref{eq:hor_vec}—\eqref{eq:vert_vec}:
\begin{equation}
    \left\{q^\mu,\pi_i\right\}=V^\mu_i\label{eq:pb_pos_mom}
\end{equation}
The intuition behind the Poisson structure \eqref{eq:pb1}—\eqref{eq:pb3} and \eqref{eq:pb_pos_mom} can be understood as follows. The body frame momenta $\pi_a$ are the infinitesimal generators of translations along the body frame axes, while $s$ is the generator of infinitesimal rotations of the body. In flat space, these generators satisfy the same commutation relations (with a minus sign) as the Lie algebra $\mathfrak{se}(2)$ of the Lie group $\mathrm{SE}(2)$ of translations and rotations in the Euclidean plane \cite{Holm2009}. However, on a curved surface, infinitesimal translations no longer commute: parallel transport of a frame along an infinitesimal loop results in the net rotation of the frame, where the rotation angle is proportional to the local Gaussian curvature of the surface. The Poisson brackets \eqref{eq:pb_pos_mom} follow from the fact that infinitesimal parallel transport and rotation of the frame are generated by the vector fields $V_i$ and the rate of change of generalized coordinates $q^j$ along such infinitesimal motions are given by the components $V_i^j$ of the corresponding vector fields.

For free motion $U(q)=0$, Hamilton's equations of motion in terms of Poisson brackets are:
\begin{eqnarray}
    \dot\pi_i&=&\left\{\pi_i,H\right\},\label{eq:ham_pb_mom}\\
    \dot q^\mu&=&\left\{q^\mu,H\right\}.\label{eq:ham_pb_pos}
\end{eqnarray}
These read explicitly as:
\begin{eqnarray}
    \dot{\pi}_a&=&\left(\frac{1}{J}-\frac{\kappa}{m}\right)\varepsilon^b_as\pi_b\label{eq:ham_lin_mom},\\
    \dot{s}&=&0\label{eq:ham_ang_mom},\\
    \dot{x}^\alpha&=&v^af_a^\alpha,\label{eq:ham_pos}\\
    \dot\psi&=&\omega-v^af_a^\alpha\Gamma_\alpha.\label{eq:ham_ang}
\end{eqnarray}
Equations \eqref{eq:ham_pos} and \eqref{eq:ham_ang} are the same as the kinematic equations of motion \eqref{eq:intr_kin_tr}—\eqref{eq:intr_kin_rot}, with $v^a$ and $\omega$ expressed in terms of the momenta using \eqref{eq:body_lin_mom_def} and \eqref{eq:body_ang_mom_def}. Equation \eqref{eq:ham_ang_mom} expresses angular momentum conservation, which follows from the fact that the Lagrangian \eqref{eq:lagr_intr} does not depend explicitly on the angle $\psi$, hence the corresponding momentum is conserved.

Equation \eqref{eq:ham_lin_mom} describes the rate of change of the body frame momenta. Note that the magnitude of the momentum $\pi_a\pi_b\delta^{ab}$ is conserved, only its direction is rotated: this is because energy is conserved (as the Hamiltonian does not depend explicitly on time), and since $s$ is conserved, so is the magnitude of the momentum. The first term on the right-hand side of \eqref{eq:ham_pb_mom} is a co-rotational derivative, originating from the fact that the $\pi_a$-s are the components of the momentum in a rotating frame, while the second term is a contribution from surface curvature, which acts as a gyroscopic force that rotates the linear momentum of the inclusion depending on its angular velocity and the local Gaussian curvature. We also mention that similar equations have been studied in the context of rotating test bodies in general relativity \cite{Kunzle1972}.

It is also instructive to write the free dynamical equations \eqref{eq:ham_lin_mom}—\eqref{eq:ham_ang_mom} extrinsically in terms of the translational velocity vector $\mathbf{v}=v^1\mathbf{f}_1+v^2\mathbf{f}_2$ and the angular velocity vector $\boldsymbol{\omega}=\omega\mathbf{f}_3$. Putting together \eqref{eq:ham_lin_mom}—\eqref{eq:ham_ang_mom} and \eqref{eq:intr_kin_tr}—\eqref{eq:intr_kin_rot} we find:
\begin{eqnarray}
	m\frac{D\mathbf{v}}{dt}&=&\kappa J\left(\boldsymbol{\omega}\times\mathbf{v}\right),\label{eq:lagr_tr_extr}\\
	J\frac{d\omega}{dt}&=&0\label{eq:lagr_rot_extr}.
\end{eqnarray}

Equation \eqref{eq:lagr_rot_extr} represents angular momentum conservation, while equation \eqref{eq:lagr_tr_extr} implies that on a curved surface, rotation and translation are coupled, resulting in a gyroscopic force in the equations of motion that rotates the direction of the velocity based on the local Gaussian curvature and the angular velocity.
\subsection{Non-conservative dynamics and integrability conditions}

We will now generalize Hamilton's equations \eqref{eq:ham_pb_mom}—\eqref{eq:ham_pb_pos} to include external forces and damping, still in a determinstic setting. We are going to add linear memoryless friction and forcing to the right-hand side of \eqref{eq:ham_pb_mom} while keeping the kinematics \eqref{eq:ham_pb_pos} unchanged:
\begin{eqnarray}
	\dot\pi_i&=&\left\{\pi_i,H\right\}-\gamma_{ij}(q)v^j+F_i(q),\label{eq:ham_pb_mom_nonc}\\
	\dot q^\mu&=&\left\{q^\mu,H\right\}.\label{eq:ham_pb_pos_nonc}
\end{eqnarray}

In equation \eqref{eq:ham_pb_mom_nonc}, $\gamma_{ij}(q)$ denotes the positive definite friction tensor (which may be anisotropic and could also depend on position) and $F_i(q)$ are conjugate forces, assumed to be independent of momenta (using the convention that $F_3(q)\equiv M(q)$ is an external torque). The velocities $v^i$ in \eqref{eq:ham_pb_mom_nonc} should be expressed in terms of the momenta $\pi_i$ via the Legendre transformation. The reason for writing the friction in terms of the velocities is that the damping originates from a quadratic Rayleigh dissipation in the \textit{velocities} added to the right-hand side of the Euler-Lagrange equations. 

The conservative forces coming from the potential term $U(q)$ in the Hamiltonian \eqref{eq:ham_quad} are given by
\begin{equation}
	F_i^{(c)}(q)=\{\pi_i,U(q)\}=\{\pi_i,q^{\mu}\}\frac{\partial U}{\partial q^\mu}=-V_i^\mu\frac{\partial{U}}{\partial q^\mu}.\label{eq:cons_force_field}
\end{equation}The work done by the forces $F_i^{(c)}$ is zero along any closed loop in the configuration space.

Since we can include conservative forces in the first term on the right-hand side of \eqref{eq:ham_pb_mom_nonc}, it is natural to reserve the last term $F_i(q)$ for non-conservative, e.g. active or "self-propulsion" forces (these are called "circulatory forces" in \cite{Ziegler1977}). In order to make sure that a given force field does not derive from a potential, we have to check that the work done by the conjugate forces is zero along each closed loop in configuration space. Restricting one's attention to infinitesimal loops results in the celebrated potential or curl conditions: a necessary condition for a force field in Euclidean space to be conservative is that its curl has to vanish.

We can generalize the curl conditions to our setting as follows. Suppose that a given force field $F_i(q)$ derives from a potential $U(q)$ as in \eqref{eq:cons_force_field}. Using the Jacobi identity for the Poisson bracket we get:
\begin{eqnarray}
	\{\pi_i,F_j(q)\}&=&\{\pi_i,\{\pi_j,U(q)\}\}\nonumber\\
	&=&\{\pi_j,\{\pi_i,U(q)\}\}+\{\{\pi_i,\pi_j\},U(q)\}\nonumber\\
	&=&\{\pi_j,F_i(q)\}+\left\{\Omega^k_{ij}(q)\pi_k,U(q)\right\}
\end{eqnarray}Expanding the Poisson brackets and rearranging, we obtain the generalized curl conditions:
\begin{equation}
	V_j^\mu\frac{\partial F_i}{\partial q^\mu}-V_i^\mu\frac{\partial F_j}{\partial q^\mu}=-\Omega^k_{ij}F_k.\label{eq:gen_curl_cond}
\end{equation}We also emphasize that the curl conditions are independent of local coordinates, since they can be written as
\begin{equation}
	V_i\left[F_j\right]-V_j\left[F_i\right]=\Omega_{ij}^kF_k,\label{eq:curl_cond_intr}
\end{equation}
where $V_i\left[F_j\right]$ denotes the application of the vector field $V_i$ to the function $F_j$. Substituting in the values of $\Omega_{ij}^k$ from \eqref{eq:pb1}—\eqref{eq:pb3} we find:
\begin{eqnarray}
	V_1\left[F_2\right]-V_2\left[F_1\right]&=&-\kappa(x)M\label{eq:curl1}\\
	V_2\left[M\right]-V_R\left[F_2\right]&=&-F_1\label{eq:curl2}\\
	V_R\left[F_1\right]-V_1\left[M\right]&=&-F_2\label{eq:curl3}
\end{eqnarray}
On the left hand side of equation \eqref{eq:curl_cond_intr} we find a kind of antisymmetric derivative of the force field $F_i(q)$, akin to the standard curl of a vector field in Euclidean space. However, contrary to the usual curl conditions, for a conservative force field this expression is in general non-zero, which can be understood intuitively as follows. When traversing a closed loop in configuration space consisting of infinitesimal translations along the body frame vectors, the work done by the translational forces are given by the expression on the left-hand side of \eqref{eq:curl_cond_intr}. However, as we have seen before, on a curved surface there is a net rotation of the frame at the end of the process, and we need to do extra work against the external torque to rotate the body back to its original position, captured by the term on the right-hand side of \eqref{eq:curl_cond_intr}.\pagebreak\subsection{Intrinsic Langevin dynamics}
In this subsection we are going to transform the Hamiltonian equations of motion \eqref{eq:ham_pb_mom_nonc}—\eqref{eq:ham_pb_pos_nonc} by adding thermal noise to the right-hand side, turning the deterministic ODEs into stochastic differential equations (SDEs). The resulting system of SDEs
\begin{eqnarray}
    d\pi_i&=&\left(\left\{\pi_i,H\right\}-\gamma_{ij}(q)v^j+F_i(q)\right)dt\nonumber\\
    &&+\sum_{n=1}^3\sigma^n_i(q)\circ dW_n,\label{eq:ou_mom}\\
    dq^i&=&\left\{q^i,H\right\}dt,\label{eq:ou_pos}
\end{eqnarray}is a generalization of the celebrated Ornstein-Uhlenbeck process \cite{Uhlenbeck1930} for a non-Euclidean phase space. In equation \eqref{eq:ou_mom} $\sigma_i^n(q)$ governs the amplitude of the noise, with the noise index $n$ running from $1$ to $3$. The $dW_n(t)$ are standard Wiener noise increments that satisfy\begin{equation}
\left\langle dW_n(t)\right\rangle=0,\quad\left\langle dW_n(t)dW_{n'}(t')\right\rangle=\delta_{nn'}\delta\left(t-t'\right).\nonumber
\end{equation} Observe that there are no ambiguities in the injection of noise into \eqref{eq:ham_pb_mom} because the body frame momenta are defined irrespective of any coordinate system. For convenience, we summarize the indices appearing in the paper, their range and role in the table below.\begin{table}[b]
\label{table:indices}
\caption{Summary of indices appearing in the paper, their range and purpose}
\begin{ruledtabular}
	\begin{tabular}{c l l}
		\textrm{Index}&\textrm{Range}&\textrm{Purpose}\\\colrule
		$\alpha$&$1,2$&\textrm{Local coordinates on the surface}\\
		$\mu$&$1,2,3$&\textrm{Local coordinates on the frame bundle}\\
		$a$&$1,2$&\textrm{Moving frames, body frame velocities, etc.}\\
		$i$&$1,2,3$\footnote{$i=3$ always refers to a rotational quantity or object (angular velocity, angular momentum, torque etc.) To emphasize this, we use the letter $R$ interchangeably with the index $i=3$.}&\textrm{Generalized velocities, momenta, forces}\\
		$I$&$1,2,3$&\textrm{Extrinsic body frame vectors}\\
		$n$&$1,2,3$&\textrm{Wiener noise index}
	\end{tabular}
\end{ruledtabular}
\end{table}

Even though we allow $\sigma^n_i(q)$ and $\gamma_{ij}(q)$ to depend on positions $q^\mu$, the noise in \eqref{eq:ou_mom}—\eqref{eq:ou_pos} is still additive — just as for a classical Ornstein-Uhlenbeck process — therefore we are free to interpret the random terms in either the Ito or Stratonovich sense. In this paper we opt for the Stratonovich interpretation, because it allows us to treat the equations in a covariant manner, without reference to any particular coordinate system. The geometric form of the system \eqref{eq:ou_mom}—\eqref{eq:ou_pos} reads:
\begin{equation}
    dy=\left(X_H+F+K\right)dt+\sum_{n=1}^3N^n\circ dW_n\label{eq:geom_sde},
\end{equation}
where $y=(q,\pi)$ is the point in phase space describing the state of the inclusion, $X_H$ is the Hamiltonian vector field (also called the Liouvillian), defined by its action on phase space functions as
\begin{eqnarray}
	X_H[f]=\{f,H\}&=&V_i^\mu\left(\frac{\partial f}{\partial q^\mu}\frac{\partial H}{\partial \pi_i}-\frac{\partial f}{\partial\pi_i}\frac{\partial H}{\partial q^\mu}\right)\nonumber\\
	& &+\Omega^k_{ij}\pi_k\frac{\partial f}{\partial \pi_i}\frac{\partial H}{\partial \pi_j}\label{eq:ham_vec_field},
\end{eqnarray}
We call
\begin{equation}
    K=-\gamma_{ij}v^j(\pi)\frac{\partial}{\partial\pi_i}=-\lambda^j_i\pi_j\frac{\partial}{\partial\pi_i}\label{eq:diss_vec}
\end{equation}
the dissipative vector field (with the matrix of inverse momentum relaxation times defined via $\lambda^i_j=\gamma_{kj}G^{ki}$),
\begin{equation}
    N^n=\sigma(q)^n_i\frac{\partial}{\partial\pi_i}\label{eq:noise_vec}
\end{equation}
are the noise vector fields for $n=1,\ldots,3$ and
\begin{equation}
    F=F_i(q)\frac{\partial}{\partial\pi_i}\label{eq:force_vec}
\end{equation}
the generalized force vector field. The geometric formulation of the stochastic equations of motion will allow us to deduce important properties about the stochastic process in an elegant, coordinate-free fashion.\section{Fokker-Planck equation}
In this section we are going to derive the Fokker-Planck equation (FPE) corresponding to the SDE \eqref{eq:geom_sde} in coordinate-free form on phase space, and obtain the fluctuation-dissipation relation for our system from it.

The phase space of the inclusion carries a natural measure, the symplectic measure \cite{Frankel2011}, which, in canonical coordinates, reads as:
\begin{eqnarray}
    \rho&=&dp_1\wedge dp_2\wedge dp_3\wedge dq^1\wedge dq^2\wedge dq^3\label{eq:sympl_coord}\\
       &=&dp_1\wedge dp_2\wedge ds\wedge dx^1\wedge dx^2\wedge d\psi\nonumber
\end{eqnarray}
Our first aim is to find the expression of $\rho$ in terms of the body frame momenta $\pi_i$. To that end, recall that the transformation between $\pi_i$ and $p_\mu$ is linear and invertible, therefore we can write
\begin{equation}
    p_\mu=\eta_\mu^i(q)\pi_i\label{eq:inv_mom}
\end{equation}
for some matrix of coefficients $\eta_\mu^i(q)$, where, in fact,
\begin{equation}
    \eta^j=\eta^j_\mu(q)dq^\mu
\end{equation} are the dual one-forms on $FS$ to the basis afforded by the horizontal and vertical vector fields $V_i$, introduced in \eqref{eq:coftr} and \eqref{eq:cofrot}. We thus have:
\begin{equation}
    \eta^j_\mu=\begin{pmatrix}
        \Theta^1_1&\Theta^1_2&0\\
        \Theta^2_1&\Theta^2_2&0\\
        \Gamma_1&\Gamma_2&1
    \end{pmatrix}\label{eq:conn_form_coord}
\end{equation}
Substituting \eqref{eq:inv_mom} and \eqref{eq:conn_form_coord} into \eqref{eq:sympl_coord}, while using the fact that the $\Theta^a$ are orthonormal, we find:\begin{eqnarray}
    \rho&=&\left(\det \eta^j_\mu\right)d\pi_1\wedge d\pi_2\wedge d\pi_3\wedge dx^1\wedge dx^2\wedge d\psi\nonumber\\
    &=&\left(\det g\right)^\frac12d\pi_1\wedge d\pi_2\wedge d\pi_3\wedge dx^1\wedge dx^2\wedge d\psi\nonumber\\
    &=&d\pi_1\wedge d\pi_2\wedge d\pi_3\wedge\mathrm{vol}_S\wedge d\psi\label{eq:sympl_measure}
\end{eqnarray}The infinitesimal generator (also called the backward Kolmogorov operator) of the stochastic process \eqref{eq:geom_sde} is given by the following second-order linear differential operator:
\begin{equation}
    L_B=C+\frac{1}{2}\sum_{n=1}^3N^nN^n,\label{eq:back_op}
\end{equation}
where $C=X_H+F+K$ is the drift vector field of the process \cite{Pavliotis2014}.

The operator which governs the forward time evolution of the probability distribution of the system is similarly a second-order linear operator, but acting on probability measures rather than functions. Nevertheless, any sufficiently well-behaved time-dependent probability measure $\zeta$ can be represented by a function $P=P(t,y)$ as $\zeta=P(t,y)\chi$ on phase space and of time, called the probability density function, once a reference measure $\chi$ is chosen on phase space. The Fokker-Planck equation or forward Kolmogorov equation
\begin{equation}
    \frac{\partial P}{\partial t}=L_FP\label{eq:forw_fpe}
\end{equation}
governs the time evolution of $P$, where the Fokker-Planck operator $L_F$ is the adjoint of $L_B$ with respect to the following inner product
\begin{equation}
    \langle f,g\rangle_\chi = \int_{T^*FS}fg\chi,\label{eq:ip_def}
\end{equation}
defined for square-integrable functions on phase space, denoted by $L^2(\chi)$ in what follows. It is important to stress that the Fokker-Planck operator $L_F$ depends on the choice of $\chi$ through the inner product \eqref{eq:ip_def}. In most applications, the state space of the stochastic process is a trivial manifold (usually $\mathbb R^n$), on which there are global coordinates $x^1,\ldots,x^n$ and $\chi=dx^1\wedge\dots\wedge dx^n$ is implicitly chosen to obtain the standard form of the FPE. For a more in-depth discussion of the issue, see e.g. \cite{Barp2021,OByrne2024}.

For our case, a natural candidate for the reference $\chi$ is the symplectic measure $\rho$. The physical implication of this choice is that $P$ will describe the phase space density of our stochastic system, while the forward operator will be
\begin{equation}
    L_F=\lambda_i^i-\left(X_H+F+K\right)+\frac12\sum_{n=1}^3N^nN^n,\label{eq:forw_fpe_op}
\end{equation}
with the details of the calculation given in the appendix. For convenience, we let $\Lambda=\mathrm{Tr}\left[\lambda^i_j\right]=\lambda^i_i$. Using the fact that for any function $f$ on phase space we have $X_H[f]=\{f,H\}$, the explicit form of the Fokker-Planck equation \eqref{eq:forw_fpe} reads
\begin{widetext}
    \begin{equation}
        \frac{\partial P}{\partial t}+\{P,H\}=\frac{\partial}{\partial\pi_i}\left[\left(\lambda^j_i(q)\pi_j-F_i(q)\right)P\right]+\frac12\sum_{n=1}^3\sigma_i^n(q)\sigma_j^n(q)\frac{\partial^2P}{\partial\pi_i\partial\pi_j}.
    \label{eq:fpe_coord}\end{equation}
\end{widetext}
Equation \eqref{eq:fpe_coord} is the generalization of the Klein-Kramers equation to our Hamiltonian system on a curved space with orientational degrees of freedom. On the left-hand side, we have the total derivative of the distribution function $P$ along the flow on phase space generated by the Hamiltonian $H$. In the deterministic setting, the right-hand side would be zero by Liouville's theorem, therefore the Fokker-Planck equation \eqref{eq:fpe_coord} can also be interpreted as the generalization of Liouville's equation to dissipative and noisy Hamiltonian dynamics, an observation which has already made by Chandrasekhar \cite{Chandrasekhar1943}.

In thermal equilibrium and the absence of external forces, we require the Gibbs measure $e^{-\beta H}\rho$ to be a steady state solution of the Fokker-Planck equation, implying a relationship between the dissipative coefficients and the noise amplitudes known as the fluctuation-dissipation relation (FDR). In our setting, we demand that the function $P_G=e^{-\beta H}$ on phase space is annihilated by the right-hand side of \eqref{eq:fpe_coord}. This gives the equation:
\begin{equation}
    \lambda_i^i-\beta\lambda^i_kG_{ij}\pi^j\pi^k=\frac{\beta}{2}\sum_{n=1}^3\sigma_j^n\sigma_k^n\left(G^{jk}-\beta\pi^j\pi^k\right),\label{eq:fdt_alg}
\end{equation}
where the indices of $\pi_i$ are raised by the inertia metric $G_{ij}$. Since in equation \eqref{eq:fdt_alg} the momenta can take arbitrary values, using $\gamma_{kj}=\lambda^i_kG_{ij}$ we obtain:
\begin{eqnarray}
    2k_BT\gamma_{jk}&=&\sum_{n=1}^3\sigma_j^n\sigma_k^n,\label{eq:fdt1}\\
    2k_BT\lambda^i_i&=&\sum_{n=1}^3\sigma_j^n\sigma_k^nG^{jk}.\label{eq:fdt2}
\end{eqnarray}
The second condition follows from the first by taking the trace, hence \eqref{eq:fdt1} is the FDR for our system. To make contact with standard notation, let us assume for simplicity that both $\gamma_{ij}$ and $\sigma_i^n$ are diagonal:
\begin{equation}
	\gamma_{ij}=\begin{pmatrix}
		\gamma_1&0&0\\
		0&\gamma_2&0\\
		0&0&\gamma_R
	\end{pmatrix},\quad\sigma_i^n=\begin{pmatrix}
		\sigma_1&0&0\\
		0&\sigma_2&0\\
		0&0&\sigma_R
	\end{pmatrix}.\label{eq:noise_fric_diag}
\end{equation}In this case, \eqref{eq:fdt1} gives:
\begin{equation}
    2k_BT\gamma_a=\sigma_a^2,\quad 2k_BT\gamma_R=\sigma_R^2
\end{equation}
Introducing the translational $D_a$ and rotational diffusion coefficients $D_R$ as
\begin{equation}
    D_a=\frac{\sigma_a^2}{2\gamma_a^2},\quad D_R=\frac{\sigma_R^2}{2\gamma_R^2},\label{eq:diff_def}
\end{equation}
the fluctation-dissipation relation \eqref{eq:fdt1} assumes the standard form:\begin{equation}
    D_a=\frac{k_BT}{\gamma_a},\quad D_R=\frac{k_BT}{\gamma_R}.\label{eq:fdt_diag}
\end{equation}
\section{Smoluchowski limit}
\subsection{Adiabatic elimination of momenta}
In this section we are going to derive the overdamped or Smoluchowski limit of the process \eqref{eq:ou_mom}—\eqref{eq:ou_pos} by adiabatically eliminating the momenta from the equations by a standard procedure, albeit we concentrate on the backward Kolmogorov operator rather than the forward one \cite{Pavliotis2014,Hottovy2012,Gardiner2004,Wilemski1976,Murphy1972}. We assume that the FDR \eqref{eq:fdt1} holds and introduce the rescaled momenta and inertia metric as follows:\begin{equation}
    \tilde{\pi}_i=\frac{\pi_i}{\sqrt m},\quad\tilde{G}_{ij}=\frac{G_{ij}}{m}.\label{eq:resc_metr}
\end{equation}
Subsequently, $\tilde{G}^{ij}$ denotes the inverse metric in \eqref{eq:resc_metr} and indices $i,j$ are raised and lowered with respect to $\tilde{G}_{ij}$. The backward Kolmogorov operator \eqref{eq:back_op} can be split as a sum of two terms:
\begin{equation}
    L_B=\frac{1}{m}L_1+\frac{1}{\sqrt m}L_2\label{eq:back_op_split},
\end{equation}
where we have (using the definition of the friction tensor and the fluctuation-dissipation relations):
\begin{eqnarray}
    L_1=&&-\gamma_{ij}\tilde{G}^{kj}\tilde{\pi}_k\frac{\partial}{\partial\tilde{\pi}_i}+\frac12\sum_{n=1}^3\sigma_i^n\sigma_j^n\frac{\partial^2}{\partial\tilde{\pi}_i\partial\tilde{\pi}_j}\nonumber\\
    &&=\gamma_{ij}\left(-\tilde{\pi}^j\frac{\partial}{\partial\tilde{\pi}_i}+k_BT\frac{\partial^2}{\partial\tilde{\pi}_i\partial\tilde{\pi}_j}\right),
\end{eqnarray}
while
\begin{equation}
    L_2=F_i\frac{\partial}{\partial{\tilde{\pi}_i}}+\sqrt m X_H.\label{eq:op_split}
\end{equation}
For the purposes of this derivation, we include all force contributions (both conservative and non-conservative) in the first term of equation \eqref{eq:op_split} and work with a purely kinetic Hamiltonian. The operator $L_1$ admits the invariant momentum density
\begin{equation}
	\xi\left(\tilde{\pi}_i\right)=\mathcal{N}\exp\left(-\frac{\beta\tilde{G}^{ij}\tilde{\pi}_i\tilde{\pi}_j}{2}\right),
\end{equation}
where $\mathcal{N}$ is a normalization factor. In what follows, we are going to work in the function space $L^2(\xi)$ of square-integrable functions with respect to the weight $\xi\left(\tilde{\pi}_i\right)$. Assume that the solution $f\left(q^\mu,\tilde{\pi}_i,t\right)\in L^2(\xi)$ of the backward Kolmogorov equation\begin{equation}
    \frac{\partial f}{\partial t}=L_Bf\label{eq:back_eq}
\end{equation} has an asymptotic expansion of the form:
\begin{equation}
    f=f_0+\sqrt m f_1+mf_2+\dots\label{eq:pow_ser}
\end{equation}
Substituting \eqref{eq:pow_ser} into \eqref{eq:back_eq} using \eqref{eq:back_op_split} and equating powers of $\sqrt m$ we find the following hierarchy:
\begin{eqnarray}
    L_1f_0&=&0,\label{eq:hier1}\\
    L_1f_1&=&-L_2f_0,\label{eq:hier2}\\
    \frac{\partial f_0}{\partial t}&=&L_1f_2+L_2f_1.\label{eq:hier3}
\end{eqnarray}
From equation \eqref{eq:hier1} and examining the null space of the operator $L_1$ in $L^2(\xi)$ it follows that $f_0$ has to be independent of momenta, so that $f_0=f_0(q,t)$. Equation \eqref{eq:hier2} then reads:
\begin{eqnarray}
    L_1f_1&=&-\sqrt m\{f_0,H\}=-\sqrt m V_i\left[f_0\right]\pi_jG^{ij}\\
    &=&-\tilde{\pi}^iV_i[f_0].
\end{eqnarray}
The general solution to \eqref{eq:hier2} in $L^2(\xi)$ is therefore given by:
\begin{equation}
    f_1=\left(\gamma^{-1}\right)^{ij}\tilde{\pi}_iV_j\left[f_0\right]+g_0(q,t),\label{eq:f1sol}
\end{equation}
where we have assumed that the friction tensor $\gamma_{ij}$ is invertible with inverse $\left(\gamma^{-1}\right)^{ij}$ and $g_0(q,t)$ is an integration constant which, in this case, is an arbitrary function of positions and time. Substituting \eqref{eq:f1sol} into equation \eqref{eq:hier3} we obtain:
\begin{eqnarray}
    \frac{\partial f_0}{\partial t}&=&L_1f_2+\left(\gamma^{-1}\right)^{ij}F_iV_j[f_0]\nonumber\\
    & &+\left\{\left(\gamma^{-1}\right)^{ij}\pi_iV_j[f_0],H\right\}+\tilde{\pi}^iV_i\left[g_0\right]
\end{eqnarray}
Expanding the Poisson bracket yields:
\begin{eqnarray}
    \frac{\partial f_0}{\partial t}&=&L_1f_2+\tilde{\pi}^iV_i\left[g_0\right]+\left(\gamma^{-1}\right)^{ij}F_iV_j[f_0]\nonumber\\
& &+\tilde{G}^{kl}\tilde{\pi}_i\tilde{\pi}_lV_k\left[\left(\gamma^{-1}\right)^{ij}V_j\left[f_0\right]\right]\nonumber\\
& &+\Omega^k_{il}\tilde{G}^{lr}\left(\gamma^{-1}\right)^{ij}\tilde{\pi}_k\tilde{\pi}_rV_j\left[f_0\right]
\label{eq:pb_exp}\end{eqnarray}
To eliminate $f_2$ and $g_0$ from equation \eqref{eq:pb_exp} we first multiply both sides by the 
the invariant density $\xi$ then integrate over momenta. The term containing $f_2$ will vanish as
\begin{equation}
	\int_{T^*FS} \xi L_1f_2\rho=\int_{T^*FS} L_1^\dagger\xi f_2\rho=0,
\end{equation}
since $L_1^\dagger$ (adjoint taken with respect to $\rho$) annihilates the stationary momentum density $\xi$, while the term containing $g_0$ is odd in momenta and will also vanish upon integration. Hence we find the limiting form of the backward equation for $f_0$:
\begin{equation}
    \frac{\partial f_0}{\partial t}=\left(F_i+\frac{\Omega^k_{ik}}{\beta}\right)\left(\gamma^{-1}\right)^{ij}V_j[f_0]+V_i\left[\frac{\left(\gamma^{-1}\right)^{ij}}{\beta}V_j\left[f_0\right]\right].\label{eq:limit_back}
\end{equation}
The second term in the parentheses in \eqref{eq:limit_back} — not anticipated by a naive overdamped limit where all inertia terms are neglected — vanishes owing to the structure of the coefficients $\Omega^{k}_{ij}$. Let us introduce the inverse noise amplitude coefficients $\Sigma(q)^i_n$ via the relations
\begin{equation}
    \Sigma(q)^i_{n'}\sigma(q)_i^{n}=\delta^n_{n'},\quad\sum_{n=1}^3\Sigma(q)^i_n\sigma(q)^n_j=\delta^i_j.
\end{equation}
Using the fluctuation-dissipation relation \eqref{eq:fdt1} we have:
\begin{equation}
    \left(\gamma^{-1}\right)^{ij}=2k_BT\sum_{n=1}^3\Sigma^i_n\Sigma^j_n.
\end{equation}
Let us define the noise vector fields $Y_n$ as
\begin{equation}
    Y_n=2k_BT\Sigma (q)^i_nV_i,
\end{equation}
which allows us to write the infinitesimal generator $\Xi$ of the limiting process as:
\begin{eqnarray}
    \Xi&=&\left(\left(\gamma^{-1}\right)^{ij}F_i+2\left(k_BT\right)^2\sum_{n=1}^3\Sigma^j_nV_i\left[\Sigma^i_n\right]\right)V_j\nonumber\\
    & &+\frac12\sum_{n=1}^3Y_nY_n.\label{eq:smolu_back_op}
\end{eqnarray}
\subsection{Intrinsic Brownian dynamics}
Using the correspondence between the infinitesimal generator of stochastic processes and stochastic differential equations \cite{Pavliotis2014} we are now in a position to write down the overdamped limit of the stochastic equations of motion:
\begin{eqnarray}
    dq^\mu&=&\left(\left(\gamma^{-1}\right)^{ij}F_i+2\left(k_BT\right)^2\sum_{n=1}^3\Sigma^j_nV_i\left[\Sigma^i_n\right]\right)V_j^\mu dt\nonumber\\
    & &+2k_BT\sum_{n=1}^3\Sigma^i_nV_i^\mu\circ dW^n.\label{eq:overdamped_stoch}
\end{eqnarray}
Note the presence of the extra drift term coming from the position dependence of the friction and diffusion tensors. To simplify the exposition, in what follows we will assume that the $\Sigma^i_n$ are independent of position, but our results can be readily extended for the case of state-dependent friction and diffusivity.

For diagonal noise and friction tensors \eqref{eq:noise_fric_diag} obeying the FDT, in local coordinates the system \eqref{eq:overdamped_stoch} takes the following familiar form:\begin{eqnarray}
    dx^\alpha&=&\sum_{a=1}^2\frac{F_af_a^\alpha}{\gamma_a}dt+\sum_{a=1}^2\sqrt{2D_a}f_a^\alpha\circ dW_a\label{eq:overdamped_sde_tr}\\
    d\psi&=&-\sum_{a=1}^2f_a^\alpha\Gamma_\alpha\left(\frac{F_a}{\gamma_a}dt+\sqrt{2D_a}\circ dW_a\right)\nonumber\\
    & &+\frac{M}{\gamma_R}dt+\sqrt{2D_R}\circ dW_R\label{eq:overdamped_sde_rot}
\end{eqnarray}
Equation \eqref{eq:overdamped_sde_tr} governs the translational motion in the overdamped limit. The first term on the right-hand side describes the effect of forces resulting in a deterministic drift. In the active Brownian particle model, $F_a$ is taken to be constant, and if $\gamma_a$ is also constant then $u_a=\frac{F_a}{\gamma_a}$ becomes a constant swimming velocity with which the particle translates in its body frame. Observe that the translational noise, which is injected in the body frame, is always multiplicative as $f_a^\alpha$ generally depends on $x^\alpha$ as well as $\psi$, therefore it is essential to fix a noise interpretation (Stratonovich in our case). Equation \eqref{eq:overdamped_sde_rot} governs the rotation of the frame, comprised of three contributions: one from the covariant derivative (the first term on the right-hand side), one from the torque and one from rotational Wiener noise. The intuitive interpretation is that the covariant derivative of the frame is given by sum of rotation from external torques and thermal noise, consistent with previous proposals \cite{Fily2016, Apaza2018, CastroVillarreal2018}.\pagebreak
\subsection{Smoluchowski equation}
In this section, we derive the Fokker-Planck or Smoluchowski equation corresponding to the stochastic process \eqref{eq:overdamped_stoch}. As discussed previously, the form of the FPE depends on the choice of reference measure. In our setting, it is natural to work with the Riemannian volume element 
\begin{equation}
    \mu_R=\mathrm{vol}_S\wedge d\psi=\eta^1\wedge\eta^2\wedge\eta^3\label{eq:riem_vol_meas}
\end{equation}
on $FS$, also appearing in the symplectic volume element in \eqref{eq:sympl_measure}. The adjoint $\Xi_F$ of the infinitesimal generator $\Xi$ with respect to the inner product $L^2\left(\mu_R\right)$ on $FS$ is given by:
\begin{equation}
    \Xi_F[P]=-\left(\gamma^{-1}\right)^{ij}\left(V_i\left[F_jP\right]-k_BTV_i\left[V_j[P]\right]\right)\label{eq:smolu_forw_op}
\end{equation}
The Smoluchowski equation reads:
\begin{equation}
    \frac{\partial P}{\partial t}=-\left(\gamma^{-1}\right)^{ij}\left(V_i\left[F_jP\right]-k_BTV_i\left[V_j[P]\right]\right)\label{eq:smolu}
\end{equation}
For this special case when the friction and diffusion tensors are independent of position we recover Brownian motion on the frame bundle $FS$ with respect to a metric whose orthonormal vector fields are given by a scaled versions $\sqrt{2D_a}V_a$ and $\sqrt{2D_R}V_R$ of the horizontal and vertical vector fields, together with a drift vector field $\left(\gamma^{-1}\right)^{ij}F_jV_i$.

This is consistent with the more general result shown in \cite{Birrell2016}, where the authors proved that the small mass limit of stochastic geodesic equations on the tangent bundle of general manifolds perturbed by linear state-independent friction and white noise is Brownian motion generated by the Laplace-Beltrami operator of the metric \cite{Hsu2002}. In our case, the extremal trajectories of the Lagrangian \eqref{eq:lagr_intr} are in fact geodesics of a metric on $FS$ whose orthonormal vector fields are $V_i$ (up to scale), and the stochastic equations of motion \eqref{eq:ou_mom}—\eqref{eq:ou_pos} are stochastic geodesic equations.
\section{Rigid inclusion on a sphere}
\subsection{Kinematics in local coordinates}
As a concrete example, we compute explicitly the equations of motion for the particular case of the sphere $S^2\subset \mathbb{R}^3$ of radius $R$ embedded in three-dimensional space as $S^2=\left\{(x,y,z)|x^2+y^2+z^2=R^2\right\}$. We use spherical polar coordinates $(\vartheta,\varphi)\in (0,\pi)\times [0,2\pi)$ on a coordinate patch excluding the north and south poles as follows:\begin{eqnarray}
	x&=&R\sin\vartheta\cos\varphi,\\
	y&=&R\sin\vartheta\sin\varphi,\\
	z&=&R\cos\vartheta.
\end{eqnarray}The restriction of the Euclidean metric to the sphere is given by $g=R^2\left(d\vartheta^2+\sin^2\vartheta d\varphi^2\right)$. It follows that the vector fields 
\begin{equation}
	\mathbf{e}_1=\frac{1}{R}\frac{\partial}{\partial\vartheta},\quad\mathbf{e}_2=\frac{1}{R\sin\vartheta}\frac{\partial}{\partial\varphi}
\end{equation}
form an orthonormal moving frame away from the poles, with the corresponding coframes
\begin{equation}
	\theta^1=Rd\vartheta,\quad \theta^2=R\sin\vartheta d\varphi.
\end{equation}
The connection form $\Gamma$ can be computed by calculating the covariant derivative of the frame vectors $\mathbf{e}_a$ or using the structure equations \eqref{eq:str_th_1}—\eqref{eq:str_th_2}. The result is\begin{equation}
	\Gamma=\cos\vartheta d\varphi=\frac{\cot \vartheta}{R}\theta^2.
\end{equation}The Gaussian curvature of the sphere is constant and equal to $\kappa=\frac{1}{R^2}$. This can also be verified using equation \eqref{eq:curv_conn_form}:
\begin{equation*}d\Gamma=-\sin\vartheta d\vartheta\wedge d\varphi=-\frac{1}{R^2}\theta^1\wedge\theta^2=-\frac{1}{R^2}\mathrm{vol}_S.
\end{equation*}With respect to our choice of moving frame, any frame field $\mathbf{f}_a(x,\psi)$ can be written as:\begin{eqnarray}
\mathbf{f}_1&=&\mathbf{e}_1\cos\psi+\mathbf{e}_2\sin\psi=\frac{\cos\psi}{R}\frac{\partial}{\partial\vartheta}+\frac{\sin\psi}{R\sin\vartheta}\frac{\partial}{\partial\varphi},\nonumber\\
\mathbf{f}_2&=&\mathbf{e}_2\cos\psi-\mathbf{e}_1\sin\psi=\frac{\cos\psi}{R\sin\vartheta}\frac{\partial}{\partial\varphi}-\frac{\sin\psi}{R}\frac{\partial}{\partial\vartheta}.\nonumber
\end{eqnarray}The horizontal \eqref{eq:hor_vec} and vertical \eqref{eq:vert_vec} vector fields are given by:
\begin{eqnarray}
V_1&=&\frac{\cos\psi}{R}\frac{\partial}{\partial\vartheta}+\frac{\sin\psi}{R\sin\vartheta}\frac{\partial}{\partial\varphi}-\frac{\sin\psi}{R\tan\vartheta}\frac{\partial}{\partial\psi},\label{eq:hor1}\\
V_2&=&\frac{\cos\psi}{R\sin\vartheta}\frac{\partial}{\partial\varphi}-\frac{\sin\psi}{R}\frac{\partial}{\partial\vartheta}-\frac{\cos\psi}{R\tan\vartheta}\frac{\partial}{\partial\psi},\label{eq:hor2}\\
V_R&=&\frac{\partial}{\partial\psi}.\label{eq:vert}
\end{eqnarray}Given the translational $v^1(t),v^2(t)$ and angular velocity $\omega(t)$ of the inclusion, its position can be integrated via:\begin{eqnarray}
\frac{d\vartheta}{dt}&=&\frac{v^1(t)\cos\psi-v^2(t)\sin\psi}{R}\\
\frac{d\varphi}{dt}&=&\frac{v^1(t)\sin\psi+v^2(t)\cos\psi}{R\sin\vartheta}\\
\frac{d\psi}{dt}&=&\omega(t)-\frac{v^1(t)\sin\psi+v^2(t)\cos\psi}{R\tan\vartheta}
\end{eqnarray}\pagebreak
\subsection{Dynamics in local coordinates}
In local coordinates, the Lagrangian \eqref{eq:lagr_coord} and Hamiltonian \eqref{eq:ham_coord} read:\begin{eqnarray}
	L&=&\frac12mR^2\left(\dot{\vartheta}^2+\sin^2\vartheta\dot{\varphi}^2\right)+\frac12J\left(\dot\psi+\cos\vartheta\dot\varphi\right)^2,\nonumber\\
	H&=&\frac{p_\vartheta^2}{2mR^2}+\frac{\left(p_\varphi-s\cos\vartheta\right)^2}{2mR^2\sin^2\vartheta}+\frac{s^2}{2J}.\label{eq:kin_ham_sphere}
\end{eqnarray}
the underdamped stochastic equations of motion \eqref{eq:ou_mom}—\eqref{eq:ou_pos} are given by (assuming that the noise and friction tensors are diagonal as in \eqref{eq:noise_fric_diag}):
\begin{eqnarray}
	d\pi_1&=&\left(\left(\frac{1}{J}-\frac{1}{mR^2}\right)s\pi_2-\frac{\gamma_1}{m}\pi_1+F_1\right)dt\nonumber\\
	& &+\gamma_1\sqrt{2D_1}\circ dW_1\label{eq:underdamped_sde_sphere_first}\\
	d\pi_2&=&\left(-\left(\frac{1}{J}-\frac{1}{mR^2}\right)s\pi_1-\frac{\gamma_2}{m}\pi_2+F_2\right)dt\nonumber\\
	& &+\gamma_2\sqrt{2D_2}\circ dW_2\\
	ds&=&-\left(\frac{\gamma_R}{J}s+M\right)dt+\gamma_R\sqrt{2D_R}\circ dW_R\\
	d\vartheta&=&\frac{\pi_1\cos\psi-\pi_2\sin\psi}{mR}dt\\
	d\varphi&=&\frac{\pi_1\sin\psi+\pi_2\cos\psi}{mR\sin\vartheta}dt\\
	d\psi&=&\left(\frac{s}{J}-\frac{\pi_1\sin\psi+\pi_2\cos\psi}{mR\tan\vartheta}\right)dt\label{eq:underdamped_sde_sphere_last}
\end{eqnarray}Further assuming that the friction and noise tensors are constant, the overdamped stochastic equations of motion \eqref{eq:overdamped_sde_tr}—\eqref{eq:overdamped_sde_rot} on the sphere read:\begin{eqnarray}
	d\vartheta&=&\left(\frac{F_1\cos\psi}{R\gamma_1}-\frac{F_2\sin\psi}{R\gamma_2}\right)dt\nonumber\\
	& &+\frac{\sqrt{2D_1}\cos\psi}{R}\circ dW_1-\frac{\sqrt{2D_2}\sin\psi}{R}\circ dW_2,\nonumber\\
	d\varphi&=&\left(\frac{F_1\sin\psi}{R\gamma_1\sin\vartheta}+\frac{F_2\cos\psi}{R\gamma_2\sin\vartheta}\right)dt\nonumber\\
	& &+\frac{\sqrt{2D_1}\sin\psi}{R\sin\vartheta}\circ dW_1+\frac{\sqrt{2D_2}\cos\psi}{R\sin\vartheta}\circ dW_2,\nonumber\\
	d\psi&=&\left(\frac{M}{\gamma_R}-\frac{\cot\vartheta}{R}\left(\frac{F_1\sin\psi}{\gamma_1}+\frac{F_2\cos\psi}{\gamma_2}\right)\right)dt\nonumber\\
	& &-\frac{\cot\vartheta}{R}\left(\sqrt{2D_1}\sin\psi\circ dW_1+\sqrt{2D_2}\cos\psi\circ dW_2\right)\nonumber\\
	& &+\sqrt{2D_R}\circ dW_R.\nonumber
\end{eqnarray}
\subsection{The Fokker-Planck and Smoluchowski equations}
The Fokker-Planck equation \eqref{eq:fpe_coord} for the phase space density $P(\vartheta,\varphi,\psi,\pi_1,\pi_2,s)\sin\vartheta d\pi_1\wedge d\pi_2\wedge ds\wedge d\vartheta\wedge d\varphi\wedge d\psi$ of the inclusion can be obtained by using the local coordinate expressions of the vector fields \eqref{eq:hor1}—\eqref{eq:vert}. Using the purely kinetic Hamiltonian \eqref{eq:ham_quad} and \eqref{eq:diff_def} we find:\begin{widetext}
	\begin{eqnarray}
		\frac{\partial P}{\partial t}&+&\frac{\pi_1\cos\psi-\pi_2\sin\psi}{mR}\frac{\partial P}{\partial\vartheta}+\frac{\pi_1\sin\psi+\pi_2\cos\psi}{mR\sin\vartheta}\frac{\partial P}{\partial\varphi}+\left(\frac{s}{J}-\frac{\pi_1\sin\psi+\pi_2\cos\psi}{mR\tan\vartheta}\right)\frac{\partial P}{\partial \psi}=\nonumber\\
		&=&\sum_{a=1}^2\frac{\partial}{\partial\pi_a}\left(\left(\frac{\gamma_a}{m}\pi_a-F_a\right)P\right)+\frac{\partial}{\partial s}\left(\left(\frac{\gamma_R}{J}s-M\right)P\right)+\sum_{a=1}^2\gamma_a^2D_a\frac{\partial^2P}{\partial \pi_a^2}+\gamma_R^2D_R\frac{\partial^2 P}{\partial s^2}.
	\end{eqnarray}
\end{widetext}
The Smoluchowski equation \eqref{eq:smolu} for the configuration space density $P(\vartheta,\varphi,\psi)\sin\vartheta d\vartheta\wedge d\varphi\wedge d\psi$ can also be obtained by putting together \eqref{eq:hor1}—\eqref{eq:vert} and \eqref{eq:smolu}. In the case of constant diagonal and friction tensors, the result is:
\begin{widetext}
	\begin{eqnarray}
		\frac{\partial P}{\partial t}&=&-\frac{\cos\psi}{\gamma_1R}\frac{\partial\left(F_1P\right)}{\partial\vartheta}-\frac{\sin\psi}{\gamma_1R\sin\vartheta}\frac{\partial\left(F_1P\right)}{\partial\varphi}+\frac{\sin\psi}{\gamma_1R\tan\vartheta}\frac{\partial\left(F_1P\right)}{\partial\psi}\nonumber\\
		& &-\frac{\cos\psi}{\gamma_2R\sin\vartheta}\frac{\partial\left(F_2P\right)}{\partial\varphi}+\frac{\sin\psi}{\gamma_2R}\frac{\partial\left(F_2P\right)}{\partial\vartheta}+\frac{\cos\psi}{\gamma_2R\tan\vartheta}\frac{\partial\left(F_2P\right)}{\partial\psi}-\frac{1}{\gamma_R}\frac{\partial\left(MP\right)}{\partial\psi}\nonumber\\
		& &+D_1\left(\frac{\cos\psi}{R}\frac{\partial}{\partial\vartheta}+\frac{\sin\psi}{R\sin\vartheta}\frac{\partial}{\partial\varphi}-\frac{\sin\psi}{R\tan\vartheta}\frac{\partial}{\partial\psi}\right)\left(\frac{\cos\psi}{R}\frac{\partial P}{\partial\vartheta}+\frac{\sin\psi}{R\sin\vartheta}\frac{\partial P}{\partial\varphi}-\frac{\sin\psi}{R\tan\vartheta}\frac{\partial P}{\partial\psi}\right)\nonumber\\
		& &+D_2\left(\frac{\cos\psi}{R\sin\vartheta}\frac{\partial}{\partial\varphi}-\frac{\sin\psi}{R}\frac{\partial}{\partial\vartheta}-\frac{\cos\psi}{R\tan\vartheta}\frac{\partial}{\partial\psi}\right)\left(\frac{\cos\psi}{R\sin\vartheta}\frac{\partial P}{\partial\varphi}-\frac{\sin\psi}{R}\frac{\partial P}{\partial\vartheta}-\frac{\cos\psi}{R\tan\vartheta}\frac{\partial P}{\partial\psi}\right)+D_R\frac{\partial^2P}{\partial\psi^2}\label{eq:smolu_coord_sphere}
	\end{eqnarray}
\end{widetext}
In the case of isotropic diffusion $D_1=D_2\equiv D$ and no drift, for the spherical part we recover the standard form of the Smoluchowski equation on the sphere \cite{Risken1996}:\begin{equation}
	\frac{\partial P}{\partial t}=\frac{D}{R^2}\left(\cot\vartheta\frac{\partial P}{\partial \vartheta}+\frac{\partial^2 P}{\partial\vartheta^2}+\frac{\partial^2 P}{\partial\varphi^2}\right)+\dots,\label{smolu_coord_sphere_base}
\end{equation}
where the omitted terms contain derivatives with respect to the orientational coordinate $\psi$.
\subsection{Active Brownian particle on the sphere}
So far the forces $F_a$ and torque $M$ have been arbitrary functions of position and orientation. To obtain a generalization of the standard active Brownian particle model \cite{Romanczuk2012} to a rigid inclusion constrained to the surface of the sphere, we make the choice\begin{equation}
	F_1=\gamma_1v_S\equiv\text{const.},\quad F_2=M=0,\label{eq:abp_law}
\end{equation}
where we assume that $\gamma_1$ is constant and we have introduced the so-called swimming velocity $v_S$ of the inclusion. The physical interpretation of this constitutive force and torque law is that the inclusion self-propels with a constant force along a fixed direction in the body frame. Using results from Section II, we can easily see that this constitutive force and torque law does not derive from a potential, as in \eqref{eq:curl2} the left-hand side is zero while the right-hand side is not. We stress that this conclusion remains valid for an arbitrary surface because the integrability conditions \eqref{eq:curl1}—\eqref{eq:curl3} hold for general surfaces. With the choice \eqref{eq:abp_law}, the overdamped stochastic equations of motion simplify to:
\begin{eqnarray}
	d\vartheta&=&\frac{v_S\cos\psi}{R}dt\nonumber\\
	& &+\frac{\sqrt{2D_1}\cos\psi}{R}\circ dW_1-\frac{\sqrt{2D_2}\sin\psi}{R}\circ dW_2,\nonumber\\
	d\varphi&=&\frac{v_S\sin\psi}{R\sin\vartheta}dt\nonumber\\
	& &+\frac{\sqrt{2D_1}\sin\psi}{R\sin\vartheta}\circ dW_1+\frac{\sqrt{2D_2}\cos\psi}{R\sin\vartheta}\circ dW_2,\nonumber\\
	d\psi&=&-\frac{v_S\cot\vartheta\sin\psi }{R}dt+\sqrt{2D_R}\circ dW_R\nonumber\\
	& &-\frac{\cot\vartheta}{R}\left(\sqrt{2D_1}\sin\psi\circ dW_1+\sqrt{2D_2}\cos\psi\circ dW_2\right).\nonumber
\end{eqnarray}
Alternatively, one can also study the corresponding Smoluchowski equation \eqref{eq:smolu_coord_sphere}. When translational noise is neglected (i.e. $D_1=D_2=0$), we recover the previous proposal of an active swimmer on the sphere in \cite{CastroVillarreal2018}.
\section{Conclusion}
In this paper the theory of intrinsic Langevin dynamics of rigid inclusions has been formulated. Starting from the kinematics of a rigid oriented inclusion constrained on a curved surface, the deterministic inertial equations of motion have been derived in Hamiltonian form from a standard Lagrangian via the method of moving frames, uncovering an interesting Poisson structure along the way. Linear coordinate-dependent friction together with general forces and torques have been added to Hamilton's equation of motion and integrability conditions for them have been obtained. By injecting Gaussian white noise to Hamilton's equations, stochastic differential equations have been proposed that model the motion of the inclusion under the influence of thermal noise. The Fokker-Planck equation of the inclusion has been given in phase space in geometric form and the momenta have been adiabatically eliminated to obtain the Smoluchowski limit of the process. An example calculation in local coordinates has also been presented on the sphere.

Our work can be applied to a wide range of different problems and extended in various dimensions. The geometric structures uncovered in the equations suggest structure-preserving numerical schemes for computer simulations \cite{Hairer2006}. The intrinsic stochastic differential equations presented in our paper can readily be integrated in local coordinates using standard methods \cite{Kloeden1992} for numerical solutions of SDEs. An immediate advantage of an intrinsic method is that the solution automatically satisfies the constraints, contrary to an extrinsic solver that usually has to employ some kind of projection to achieve that. However, since any coordinate system is only locally valid on a non-trivial manifold, it is challenging to extend these methods globally, as one generally has to change charts during integration. It would therefore be desirable to develop a numerical scheme based on the method of moving frames, which could combine the benefits of both the intrinsic and extrinsic viewpoints, as it would respect the geometry of the problem but also facilitate global integration by utilizing frames in the ambient space.

It is straightforward to extend our theory to many interacting rigid inclusions on a surface by considering a multi-body Hamiltonian and an interaction potential. It would be interesting to gain further insight into how the fascinating collective effects produced by active particles such as motility-induced phase separation \cite{Cates2015,Iyer2023} or intriguing patterns in polymeric systems \cite{Saintillan2018,Manna2019} manifest on a curved surface \cite{Praetorius2018}. By a suitable coarse-graining procedure, field theories could be written down and studied over curved surfaces, where the solutions of the resulting PDEs would also be dramatically altered by the geometry and topology of the surface \cite{Fily2016,Marchetti2017,CastroVillarreal2018}. Another exciting area of future research could be the extension of the well-established linear response theory \cite{Kubo1991} to the manifold setting — it is quite likely that curvature terms would influence e.g. the Green-Kubo relations, connecting transport coefficients and correlation functions. The integrability conditions for the forces and torques and the geometric form of the Fokker-Planck equation could shed further light on questions related to irreversibility \cite{Cates2016,OByrne2024}.

Large deviations and rare events is an important branch of statistical physics \cite{Wentzell1970,Graham1973,Graham1987,Touchette2009} that has found many applications in soft and active matter. Our formalism is very well suited to study such problems in a manifold setting, by looking at for example escape and stability problems at stationary points of potentials on surfaces \cite{Berglund2013,Zhou2016}. Control of active matter \cite{Fodor2024} is also an emerging field of research that has received considerable attention lately and the investigation of optimal protocols on surfaces has already been initiated \cite{Piro2021}. We hope to extend these results with the aid of the theory developed above. Finally, we aim to generalize our setup to the intriguing and challenging scenario when there is feedback between the motion of the inclusion and the surface, as is the case for most problems in biology.
\begin{acknowledgements}We thank Prof. Dr. Tanja Schilling for helpful and interesting discussions. The first author thanks the organisers of the DPG Spring Meeting (17—22 March 2024, Berlin, Germany) for an opportunity to present this work. This research was supported by the Engineering and Physical Sciences Research Council (B.N., award no. EP/W524141/1).
\end{acknowledgements}
\bibliography{main}
\appendix
\section{Derivation of Poisson brackets}
In this section we are going to derive the Poisson bracket structure \eqref{eq:pb1}—\eqref{eq:pb_pos_mom} in terms of the body frame momenta and positional coordinates. Using \eqref{eq:inv_mom} and standard properties of Poisson brackets we find:
\begin{equation}
    \left\{q^\mu,p_\nu\right\}=\delta^\mu_\nu=\left\{q^\mu,\pi_j\eta^j_\nu(q)\right\}=\eta^j_\nu(q)\{q^\mu,\pi_j\}.
\end{equation}
Hence:
\begin{equation}
    \{q^\mu,\pi_j\}=V^\mu_j(q)
\end{equation}
Similarly, we have:
\begin{eqnarray}
    \{\pi_i,\pi_j\}&=&\left\{V_i^\mu(q)p_\mu,V_j^\nu(q)p_\nu\right\}\nonumber\\
    &=&V_j^\nu\frac{\partial V_i^\mu}{\partial q^\nu}p_\mu-V_i^\mu\frac{\partial V_j^\nu}{\partial q^\mu}p_\nu\nonumber\\
    &=&-\left[V_i,V_j\right]^\mu p_\mu\nonumber\\
    &=&-\left[V_i,V_j\right]^\mu\eta^k_\mu\pi_k\nonumber\\
    &=&\Omega^k_{ij}\pi_k
\end{eqnarray}
where square brackets denote the commutator or Lie bracket of vector fields. There are various ways to finish the calculation from this point: for example, one can compute the commutators using the coordinate expressions \eqref{eq:hor_vec}—\eqref{eq:vert_vec}. Here we simplify the calculations using the following exterior calculus identity \cite{Frankel2011}:
\begin{equation}
    d\eta^i\left(V_j,V_k\right)=V_j\left[\eta^i\left(V_k\right)\right]-V_k\left[\eta^i\left(V_j\right)\right]-\eta^i\left(\left[V_j,V_k\right]\right).
\end{equation}
As $\eta^i\left(V_j\right)=\delta^i_j$, the first two terms on the right-hand side vanish, leading to:
\begin{equation}
    d\eta^i\left(V_j,V_k\right)=-\eta^i\left(\left[V_j,V_k\right]\right)=\Omega^i_{jk}.
\end{equation}
Therefore it suffices to compute the exterior derivatives of the $\eta^i$-s to find the Poisson brackets. For that purpose, we recall the structure equations of the connection \cite{Frankel2011}:
\begin{eqnarray}
	d\theta^1-\Gamma\wedge \theta^2&=&0,\label{eq:str_th_1}\\
	d\theta^2+\Gamma\wedge\theta^1&=&0,\label{eq:str_th_2}
\end{eqnarray}
which represent the fact that the Levi-Civita connection has zero torsion. After some calculations using these equations and \eqref{eq:coftr}—\eqref{eq:cofrot}, we get:
\begin{eqnarray}
    d\eta^1&=&d\Theta^1=\eta^3\wedge\eta^2,\label{eq:struct_tr_1}\\
    d\eta^2&=&d\Theta^2=\eta^1\wedge\eta^3,\label{eq:struct_tr_2}\\
    d\eta^3&=&d\Gamma=-\kappa\theta^1\wedge\theta^2=-\kappa\eta^1\wedge\eta^2.\label{eq:struct_rot}
\end{eqnarray}
It follows that the Poisson brackets between the body frame linear and angular momenta are:
\begin{eqnarray}
    \{\pi_1,\pi_2\}&=&\Omega^i_{12}\pi_i=\Omega^3_{12}\pi_3=-\kappa s,\\
    \{\pi_2,\pi_3\}&=&\Omega^i_{23}\pi_i=\Omega^1_{23}\pi_1=-\pi_1,\\
    \{\pi_3,\pi_1\}&=&\Omega^i_{31}\pi_i=\Omega^2_{31}\pi_2=-\pi_2.
\end{eqnarray}
\section{Derivation of the forward Fokker-Planck operator}
In this section we summarize the relationship between the infinitesimal generator of a continuous and time-homogeneous Markov process $X_t$ on a manifold and the forward Kolmogorov operator in the Fokker-Planck equation, for a more detailed account, consult e.g. \cite{Pavliotis2014}.

Let $X_t$ evolve on a smooth manifold $M$ of dimension $m$, starting from the deterministic initial condition $X_0=x_0$ for some $x_0\in M$. The infinitesimal generator $L_B$ of $X_t$, also called the backward Kolmogorov operator, is a differential operator defined by the following property:
\begin{equation}\label{eq:back_op_def_abs}
    \frac{d}{dt}\bigg|_{t=0}\mathbb{E}\left[f\left(X_t\right)\right]=L_B\left(x_0\right)[f],
\end{equation}
where $f:M\to\mathbb R$ is an arbitrary smooth function on $M$. In this article, we restrict our attention to \textit{diffusion} processes, where $L_B$ is a second-order differential operator. In local coordinates $x^i$ it can always be expressed in the form:
\begin{equation}
    L_B(x)=a^i(x)\frac{\partial}{\partial x^i}+\frac12b^{ij}(x)\frac{\partial^2}{\partial x^i\partial x^j},
\end{equation}
where the $a^i(x)$ and $b^{ij}(x)=b^{ji}(x)$ are smooth functions on $M$, called the drift and diffusion coefficients, respectively. It is important to stress that under a change of coordinates, the components $b^{ij}(x)$ transform as the components of a symmetric tensor field contrary to $a^i(x)$ that do \textit{not} transform as the components of a vector field. This is related to the fact that the drift of an Ito SDE is not a vector field.

To work out the forward Kolmogorov or Fokker-Planck equation of the process, we introduce the time-dependent probability measure $p_t\chi$ on $M$ describing the distribution of $X_t$ at time $t$, where $\chi$ is an arbitrary reference measure on $M$ and $p_t:M\to\mathbb [0,1]$ is the probability density function with respect to $\chi$. For processes taking place on $\mathbb R^d$, a natural choice is the Lebesgue measure $dx^1\wedge\dots\wedge dx^d$, but, as we have seen previously, on general manifolds different choices might be more appropriate. We may write \eqref{eq:back_op_def_abs} as:
\begin{equation}\label{eq:almost_fpe}
    \int_M\frac{\partial p_t}{\partial t}f\chi=\int_ML_B[f]p_t\chi=\int_ML_B^\dagger[p_t]f\chi,
\end{equation}
where $L_B^\dagger$ denotes the adjoint of $L_B$ in $L^2(\chi)$. Since in \eqref{eq:almost_fpe} the function $f$ is arbitrary, upon localizing we obtain the forward Kolmogorov or Fokker-Planck equation
\begin{equation}
    \frac{\partial p_t}{\partial t}=L_F\left[p_t\right],
\end{equation}
where $L_F=L_B^\dagger$.

The computation of $L_F$ simplifies when the infinitesimal generator is in so-called Hörmander form, i.e. it is of the form
\begin{equation}
    L_B=V+\sum_{i=1}^nY_iY_i,
\end{equation}
for some vector fields $V$ and $Y_1,\dots , Y_n$ on $M$. This is because for any vector field $A$ and real-valued functions $f,g$ on $M$ we have:
\begin{eqnarray}
    \int_MA[f]g\chi&=&\int_MA^i(x)\frac{\partial f}{\partial x^i} g(x)\chi(x)dx^1\wedge\dots\wedge dx^m\nonumber\\
    &=&-\int_Mf(x)\frac{\partial \left(A^i(x)g(x)\chi(x)\right)}{\partial x^i}dx^1\wedge\dots\wedge dx^m\nonumber,
\end{eqnarray}
after performing an integration by parts, neglecting boundary terms. Expanding the derivative, we get:
\begin{widetext}
    \begin{eqnarray}
        \int_MA[f]g\chi=\left\langle A[f],g\right\rangle_\chi&=&-\int_Mf(x)\chi(x)\left[\frac{1}{\chi(x)}\frac{\partial\left(A^i(x)\chi(x)\right)}{\partial x^i}g(x)+A^i\frac{\partial g}{\partial x^i}\right]dx^1\wedge\dots\wedge dx^m\\
        &=&-\int_Mf\left(\left(\mathrm{div}_\chi A\right)g+A[g]\right)\chi=-\left\langle f,\left(\mathrm{div}_\chi A\right)g+A[g]\right\rangle_\chi,
    \end{eqnarray}
\end{widetext}
where we have introduced the divergence \begin{equation}
    \mathrm{div}_\chi A=\frac{1}{\chi(x)}\frac{\partial\left(A^i(x)\chi(x)\right)}{\partial x^i}
\end{equation} of the vector field $A$ with respect to the measure $\chi$. The intrinsic definition of the divergence uses the Lie derivative: since $\mathcal{L}_A\chi$ is also a volume form, it has to be proportional to the volume form $\chi$, and this factor of proportionality is the divergence of the vector field $A$ with respect to $\chi$ \cite{Frankel2011}. It follows that the $L^2(\chi)$-adjoint of the linear operator $f\mapsto A[f]$ is given by \begin{equation}
	g\mapsto -\left(\mathrm{div}_\chi A\right)g-A[g].
\end{equation}
This relation allows us to compute the adjoint of any operator in Hörmander form iteratively. For the underdamped system \eqref{eq:geom_sde}, we have:
\begin{equation}
	\mathrm{div}_\mu F=\mathrm{div}_\mu X_H=\mathrm{div}_\mu N^n=0,\quad\mathrm{div}_\mu K=-\lambda^i_i.
\end{equation}
These relations follow from a short explicit computation using \eqref{eq:sympl_measure}. The fact that $\mathrm{div}_\mu X_H=0$ is essentially the statement of Liouville's theorem, while a generic feature of dissipative vector fields is that they shrink volume in phase space, hence the negative divergence of $K$ is expected.

To compute the forward operator \eqref{eq:smolu_forw_op} of the Smoluchowski equation, we essentially have to calculate the divergence of the vector fields $V_i$ with respect to the Riemannian volume form \eqref{eq:riem_vol_meas}. This can again be done in coordinates, since the coordinate expressions of both the vector fields $V_i$ and the dual frames $\eta^j$ are known from Subsection \ref{subsec:geom_prelim}. A faster but more abstract method is to use the definition of divergence in terms of Lie derivatives and Cartan's magic formula \cite{Frankel2011}:
\begin{eqnarray}
	\mathcal{L}_{V_1}\mu_R&=&\left(\mathrm{div}_{\mu_R}V_1\right)\mu_R\nonumber\\
	&=&d\iota_{V_1}\left(\eta^1\wedge\eta^2\wedge\eta^3\right)\nonumber\\
	&=&d\left(\eta^2\wedge \eta^3\right)\nonumber\\
	&=&d\left(-d\eta^1\right)\nonumber\\
	&=&0
\end{eqnarray}
Similarly, $\mathcal{L}_{V_2}\mu_R=-d\left(\eta^1\wedge \eta^3\right)=-dd\eta^2=0$ and
\begin{eqnarray}
	\mathcal{L}_{V_3}\mu_R&=&d\left(\eta^1\wedge\eta^2\right)=-d\left(\frac{1}{\kappa}d\eta^3\right)\nonumber\\
	&=&-\frac{1}{\kappa}(\mathbf{f}_1[\kappa]\eta^1+\mathbf{f}_2[\kappa]\eta^2)\wedge\left(\eta^1\wedge\eta^2\right)\nonumber\\
	&=&0.
\end{eqnarray}
Thus all $V_i$-s are divergence-free with respect to $\mu_R$, which simplifies greatly the calculation of the forward Smoluchowski operator.
\end{document}

%% file: fig2.pdf_tex
\begingroup%
  \makeatletter%
  \providecommand\color[2][]{%
    \errmessage{(Inkscape) Color is used for the text in Inkscape, but the package 'color.sty' is not loaded}%
    \renewcommand\color[2][]{}%
  }%
  \providecommand\transparent[1]{%
    \errmessage{(Inkscape) Transparency is used (non-zero) for the text in Inkscape, but the package 'transparent.sty' is not loaded}%
    \renewcommand\transparent[1]{}%
  }%
  \providecommand\rotatebox[2]{#2}%
  \newcommand*\fsize{\dimexpr\f@size pt\relax}%
  \newcommand*\lineheight[1]{\fontsize{\fsize}{#1\fsize}\selectfont}%
  \ifx\svgwidth\undefined%
    \setlength{\unitlength}{561.68099952bp}%
    \ifx\svgscale\undefined%
      \relax%
    \else%
      \setlength{\unitlength}{\unitlength * \real{\svgscale}}%
    \fi%
  \else%
    \setlength{\unitlength}{\svgwidth}%
  \fi%
  \global\let\svgwidth\undefined%
  \global\let\svgscale\undefined%
  \makeatother%
  \begin{picture}(1,0.63954099)%
    \lineheight{1}%
    \setlength\tabcolsep{0pt}%
    \put(0,0){\includegraphics[width=\unitlength,page=1]{fig2.pdf}}%
    \put(0.85923271,0.07153539){\color[rgb]{0,0,0}\makebox(0,0)[lt]{\lineheight{1.25}\smash{\begin{tabular}[t]{l}$S$\end{tabular}}}}%
    \put(0.20204008,0.08617119){\color[rgb]{0,0,1}\makebox(0,0)[lt]{\lineheight{1.25}\smash{\begin{tabular}[t]{l}$\mathbf{e}_1$\end{tabular}}}}%
    \put(0.35495852,0.27022119){\color[rgb]{0,0,1}\makebox(0,0)[lt]{\lineheight{1.25}\smash{\begin{tabular}[t]{l}$\mathbf{e}_1$\end{tabular}}}}%
    \put(0.20204008,0.38353144){\color[rgb]{0,0,1}\makebox(0,0)[lt]{\lineheight{1.25}\smash{\begin{tabular}[t]{l}$\mathbf{e}_1$\end{tabular}}}}%
    \put(0.53812868,0.43665289){\color[rgb]{0,0,1}\makebox(0,0)[lt]{\lineheight{1.25}\smash{\begin{tabular}[t]{l}$\mathbf{e}_1$\end{tabular}}}}%
    \put(0.70230092,0.16577328){\color[rgb]{0,0,1}\makebox(0,0)[lt]{\lineheight{1.25}\smash{\begin{tabular}[t]{l}$\mathbf{e}_1$\end{tabular}}}}%
    \put(0.68338081,0.28761114){\color[rgb]{0,0,1}\makebox(0,0)[lt]{\lineheight{1.25}\smash{\begin{tabular}[t]{l}$\mathbf{e}_1$\end{tabular}}}}%
    \put(0.79298506,0.42442038){\color[rgb]{0,0,1}\makebox(0,0)[lt]{\lineheight{1.25}\smash{\begin{tabular}[t]{l}$\mathbf{e}_1$\end{tabular}}}}%
    \put(0,0){\includegraphics[width=\unitlength,page=2]{fig2.pdf}}%
    \put(0.10810529,0.18716219){\color[rgb]{0,0,1}\makebox(0,0)[lt]{\lineheight{1.25}\smash{\begin{tabular}[t]{l}$\mathbf{e}_2$\end{tabular}}}}%
    \put(0.09854271,0.49289057){\color[rgb]{0,0,1}\makebox(0,0)[lt]{\lineheight{1.25}\smash{\begin{tabular}[t]{l}$\mathbf{e}_2$\end{tabular}}}}%
    \put(0.56608639,0.57413055){\color[rgb]{0,0,1}\makebox(0,0)[lt]{\lineheight{1.25}\smash{\begin{tabular}[t]{l}$\mathbf{e}_2$\end{tabular}}}}%
    \put(0.56608643,0.23736424){\color[rgb]{0,0,1}\makebox(0,0)[lt]{\lineheight{1.25}\smash{\begin{tabular}[t]{l}$\mathbf{e}_2$\end{tabular}}}}%
    \put(0.61854816,0.34999251){\color[rgb]{0,0,1}\makebox(0,0)[lt]{\lineheight{1.25}\smash{\begin{tabular}[t]{l}$\mathbf{e}_2$\end{tabular}}}}%
    \put(0.72204554,0.49884824){\color[rgb]{0,0,1}\makebox(0,0)[lt]{\lineheight{1.25}\smash{\begin{tabular}[t]{l}$\mathbf{e}_2$\end{tabular}}}}%
    \put(0,0){\includegraphics[width=\unitlength,page=3]{fig2.pdf}}%
    \put(0.19894404,0.14956608){\color[rgb]{1,0,0}\makebox(0,0)[lt]{\lineheight{1.25}\smash{\begin{tabular}[t]{l}$\mathbf{f}_1$\end{tabular}}}}%
    \put(0.21555408,0.4422205){\color[rgb]{1,0,0}\makebox(0,0)[lt]{\lineheight{1.25}\smash{\begin{tabular}[t]{l}$\mathbf{f}_1$\end{tabular}}}}%
    \put(0.58646936,0.52063054){\color[rgb]{1,0,0}\makebox(0,0)[lt]{\lineheight{1.25}\smash{\begin{tabular}[t]{l}$\mathbf{f}_1$\end{tabular}}}}%
    \put(0.62092675,0.21948272){\color[rgb]{1,0,0}\makebox(0,0)[lt]{\lineheight{1.25}\smash{\begin{tabular}[t]{l}$\mathbf{f}_1$\end{tabular}}}}%
    \put(0.31429431,0.34652366){\color[rgb]{0,0,1}\makebox(0,0)[lt]{\lineheight{1.25}\smash{\begin{tabular}[t]{l}$\mathbf{e}_2$\end{tabular}}}}%
    \put(0.05435416,0.17881746){\color[rgb]{1,0,0}\makebox(0,0)[lt]{\lineheight{1.25}\smash{\begin{tabular}[t]{l}$\mathbf{f}_2$\end{tabular}}}}%
    \put(0.02430716,0.43142305){\color[rgb]{1,0,0}\makebox(0,0)[lt]{\lineheight{1.25}\smash{\begin{tabular}[t]{l}$\mathbf{f}_2$\end{tabular}}}}%
    \put(0.48435375,0.58850092){\color[rgb]{1,0,0}\makebox(0,0)[lt]{\lineheight{1.25}\smash{\begin{tabular}[t]{l}$\mathbf{f}_2$\end{tabular}}}}%
    \put(0.49461283,0.14322107){\color[rgb]{1,0,0}\makebox(0,0)[lt]{\lineheight{1.25}\smash{\begin{tabular}[t]{l}$\mathbf{f}_2$\end{tabular}}}}%
    \put(0.41091887,0.50322439){\color[rgb]{0,0.50196078,0}\makebox(0,0)[lt]{\lineheight{1.25}\smash{\begin{tabular}[t]{l}$x(t)$\end{tabular}}}}%
    \put(0,0){\includegraphics[width=\unitlength,page=4]{fig2.pdf}}%
    \put(0.17716147,0.1247978){\color[rgb]{0.50196078,0,0.50196078}\makebox(0,0)[lt]{\lineheight{1.25}\smash{\begin{tabular}[t]{l}$\psi$\end{tabular}}}}%
    \put(0,0){\includegraphics[width=\unitlength,page=5]{fig2.pdf}}%
    \put(0.15095887,0.37761837){\color[rgb]{0.50196078,0,0.50196078}\makebox(0,0)[lt]{\lineheight{1.25}\smash{\begin{tabular}[t]{l}$\psi$\end{tabular}}}}%
    \put(0,0){\includegraphics[width=\unitlength,page=6]{fig2.pdf}}%
    \put(0.54496504,0.4998976){\color[rgb]{0.50196078,0,0.50196078}\makebox(0,0)[lt]{\lineheight{1.25}\smash{\begin{tabular}[t]{l}$\psi$\end{tabular}}}}%
    \put(0,0){\includegraphics[width=\unitlength,page=7]{fig2.pdf}}%
    \put(0.62201224,0.19422487){\color[rgb]{0.50196078,0,0.50196078}\makebox(0,0)[lt]{\lineheight{1.25}\smash{\begin{tabular}[t]{l}$\psi$\end{tabular}}}}%
    \put(0,0){\includegraphics[width=\unitlength,page=8]{fig2.pdf}}%
  \end{picture}%
\endgroup%

%% file: fig1.pdf_tex
\begingroup%
  \makeatletter%
  \providecommand\color[2][]{%
    \errmessage{(Inkscape) Color is used for the text in Inkscape, but the package 'color.sty' is not loaded}%
    \renewcommand\color[2][]{}%
  }%
  \providecommand\transparent[1]{%
    \errmessage{(Inkscape) Transparency is used (non-zero) for the text in Inkscape, but the package 'transparent.sty' is not loaded}%
    \renewcommand\transparent[1]{}%
  }%
  \providecommand\rotatebox[2]{#2}%
  \newcommand*\fsize{\dimexpr\f@size pt\relax}%
  \newcommand*\lineheight[1]{\fontsize{\fsize}{#1\fsize}\selectfont}%
  \ifx\svgwidth\undefined%
    \setlength{\unitlength}{469.46115569bp}%
    \ifx\svgscale\undefined%
      \relax%
    \else%
      \setlength{\unitlength}{\unitlength * \real{\svgscale}}%
    \fi%
  \else%
    \setlength{\unitlength}{\svgwidth}%
  \fi%
  \global\let\svgwidth\undefined%
  \global\let\svgscale\undefined%
  \makeatother%
  \begin{picture}(1,0.80238277)%
    \lineheight{1}%
    \setlength\tabcolsep{0pt}%
    \put(0,0){\includegraphics[width=\unitlength,page=1]{fig1.pdf}}%
    \put(0.63949483,0.49733653){\color[rgb]{0,0.02352941,1}\makebox(0,0)[lt]{\lineheight{1.25}\smash{\begin{tabular}[t]{l}$\mathbf{v}$\end{tabular}}}}%
    \put(0.58367103,0.75384831){\color[rgb]{0,0,0}\makebox(0,0)[lt]{\lineheight{1.25}\smash{\begin{tabular}[t]{l}$\mathbf{f}_3=\mathbf{n}$\end{tabular}}}}%
    \put(0.5606923,0.44954976){\color[rgb]{0,0,0}\makebox(0,0)[lt]{\lineheight{1.25}\smash{\begin{tabular}[t]{l}$\mathbf{f}_1$\end{tabular}}}}%
    \put(0.66427686,0.53230442){\color[rgb]{0,0,0}\makebox(0,0)[lt]{\lineheight{1.25}\smash{\begin{tabular}[t]{l}$\mathbf{f}_2$\end{tabular}}}}%
    \put(0,0){\includegraphics[width=\unitlength,page=2]{fig1.pdf}}%
    \put(0.59525115,0.71275917){\color[rgb]{1,0,0.04313725}\makebox(0,0)[lt]{\lineheight{1.25}\smash{\begin{tabular}[t]{l}$\omega_3=\omega$\end{tabular}}}}%
    \put(0.73495135,0.0490971){\color[rgb]{0,0,0}\makebox(0,0)[lt]{\lineheight{1.25}\smash{\begin{tabular}[t]{l}$S$\end{tabular}}}}%
    \put(0.57350678,0.48587804){\color[rgb]{1,0,0}\makebox(0,0)[lt]{\lineheight{1.25}\smash{\begin{tabular}[t]{l}$\omega_1$\end{tabular}}}}%
    \put(0,0){\includegraphics[width=\unitlength,page=3]{fig1.pdf}}%
    \put(0.63939326,0.57410808){\color[rgb]{1,0,0}\makebox(0,0)[lt]{\lineheight{1.25}\smash{\begin{tabular}[t]{l}$\omega_2$\end{tabular}}}}%
    \put(0,0){\includegraphics[width=\unitlength,page=4]{fig1.pdf}}%
    \put(0.52561261,0.60495444){\color[rgb]{0,0.50196078,0}\makebox(0,0)[lt]{\lineheight{1.25}\smash{\begin{tabular}[t]{l}$x\in S$\end{tabular}}}}%
    \put(0.8889526,0.45750768){\color[rgb]{0.10196078,0.10196078,0.10196078}\makebox(0,0)[lt]{\lineheight{1.25}\smash{\begin{tabular}[t]{l}$T_xS$\end{tabular}}}}%
    \put(0,0){\includegraphics[width=\unitlength,page=5]{fig1.pdf}}%
  \end{picture}%
\endgroup%